\documentclass{pasj01}
\usepackage{lscape}
\usepackage{comment}
\usepackage{ccaption}
\usepackage {threeparttable}
\usepackage{color}
\draft 
\Received{$\langle$reception date$\rangle$}
\Accepted{$\langle$acception date$\rangle$}
\Published{$\langle$publication date$\rangle$}

\begin{document}

\title{Massive star formation in W51A triggered by cloud-cloud collisions}
\author{Shinji FUJITA$^{1,2,3*}$}
\author{Kazufumi TORII$^{4}$}%
\author{Nario KUNO$^{2,3}$}
\author{Atsushi NISHIMURA$^{1}$}%
\author{Tomofumi UMEMOTO$^{4,5}$}
\author{Tetsuhiro MINAMIDANI$^{4,5}$}%
\author{Mikito KOHNO$^{1}$}%
\author{Mitsuyoshi YAMAGISHI$^{6}$}%
\author{Tomoka TOSAKI$^{7}$}%
\author{Mitsuhiro MATSUO$^{8}$}%
\author{Yuya TSUDA$^{9}$}%
\author{Rei ENOKIYA$^{1}$}%
\author{Kengo TACHIHARA$^{1}$}%
\author{Akio OHAMA$^{1}$}%
\author{Hidetoshi SANO$^{1}$}%
\author{Kazuki OKAWA$^{1}$}%
\author{Katsuhiro HAYASHI$^{1}$}%
\author{Satoshi YOSHIIKE$^{1}$}%
\author{Daichi TSUTSUMI$^{1}$}%
\author{Yasuo FUKUI$^{1}$}%
\author{other FUGIN members}%

\altaffiltext{1}{Department of Astrophysics, Nagoya University, Furo-cho, Chikusa-ku, Nagoya, Aichi, Japan 464-8602}
\altaffiltext{2}{Department of Physics, Graduate School of Pure and Applied Sciences, University of Tsukuba, 1-1-1 Tennodai, Tsukuba, Ibaraki, Japan 305-8577}
\altaffiltext{3}{Tomonaga Center for the History of the Universe, University of Tsukuba, 1-1-1 Tennodai, Tsukuba, Ibaraki 305-8571, Japan}
\altaffiltext{4}{Nobeyama Radio Observatory, 462-2 Minamimaki, Minamisaku, Nagano, Japan 384-1305}
\altaffiltext{5}{Department of Astronomical Science, School of Physical Science, SOKENDAI (The Graduate University for Advanced Studies), 2-21-1, Osawa, Mitaka, Tokyo 181-8588, Japan}
\altaffiltext{6}{Institute of Space and Astronautical Science, Japan Aerospace Exploration Agency, Chuo-ku, Sagamihara 252-5210, Japan}
\altaffiltext{7}{Department of Geoscience, Joetsu University of Education, Joetsu, Niigata, Japan 943-8512}
\altaffiltext{8}{Graduate Schools of Science and Engineering, Kagoshima University, 1-21-35 Korimoto, Kagoshima, Kagoshima, Japan 890-0065}
\altaffiltext{9}{Meisei University, 2-1-1 Hodokubo, Hino, Tokyo, Japan 191-0042}

\email{fujita.shinji@a.phys.nagoya-u.ac.jp}



\maketitle

\begin{abstract}
W51A is one of the most active star-forming regions in the Milky Way, which includes copious amounts of molecular gas with a total mass of $\sim6\times10^5$\,$M_\odot$.
The molecular gas has multiple velocity components over $\sim20$\,km\,s$^{-1}$, and interactions between these components have been discussed as the mechanism that triggered the massive star formation in W51A.
In this paper, we report an observational study of the molecular gas in W51A using the new $^{12}$CO, $^{13}$CO, and C$^{18}$O ($J$=1--0) data covering a $1\fdg4\times1\fdg0$ area of W51A obtained with the Nobeyama 45-m telescope at 20$''$ resolution.
Our CO data resolved four discrete velocity clouds with sizes and masses of $\sim 30$\,pc and $1.0$--$1.9\times10^5\ M_{\odot}$ around radial velocities of 50, 56, 60, and 68 km s$^{-1}$. 
Toward the central part of the H{\sc ii} region complex G49.5-0.4 in W51A, in which the bright stellar clusters IRS\,1 and IRS\,2 are located, we identified four C$^{18}$O clumps having sizes of $\sim1$\,pc and column densities of higher than $10^{23}$\,cm$^{-2}$, which are each embedded within the four velocity clouds. 
These four clumps are concentrated within a small area of 5\,pc, but show a complementary distribution on the sky. 
In the position-velocity diagram, these clumps are connected with each other by bridge features having weak intensities.
The high intensity ratios of $^{13}$CO ($J$=3--2)/($J$=1--0) also indicate that these four clouds are associated with the H{\sc ii} regions including IRS\,1 and IRS\,2.
We also revealed that, in the other bright H{\sc ii} region complex G49.4-0.3, the 50, 60, and 68\,km\,s$^{-1}$ clouds show a complementary distribution, with two bridge features connecting between the 50 and 60\,km\,s$^{-1}$ clouds and the 60 and 68\,km\,s$^{-1}$ clouds.
An isolated compact H{\sc ii} region G49.57-0.27 located $\sim$15\,pc north of G49.5-0.4 also shows a complementary distribution and a bridge feature.
The complementary distribution on the sky and the broad bridge feature in the position-velocity diagram suggest collisional interactions among the four velocity clouds in W51A.
The timescales of the collisions can be estimated to be several 0.1\,Myrs as crossing times of the collisions, which are consistent with the ages of the H{\sc ii} regions measured from the sizes of the H{\sc ii} regions with the 21\,cm continuum data.
We discuss a scenario of the cloud-cloud collisions and massive star formation in W51A by comparing these with recent observational and theoretical studies of cloud-cloud collision.
\end{abstract}

\section{Introduction}

\textcolor{black}{\subsection{W51}}
Massive stars are influential in the galactic environment by releasing heavy elements and a large amount of energy in ultra-violet (UV) radiation, stellar winds, outflows, and supernova explosions. 
It is therefore of fundamental importance to understand the mechanisms of massive star formation, and considerable efforts have been made so far (e.g., \cite{1987ApJ...319..850W}; \cite{2007ARA&A..45..481Z}; \cite{2014prpl.conf..149T}). 
\textcolor{black}{Since giant molecular clouds (GMCs) are the principal sites of massive star formation (e.g., \cite{2007ARA&A..45..481Z}), performing large-scale molecular line observations on GMCs at high spatial resolution is important. }
\textcolor{black}{A} spatial resolution less than 1\,pc allows one to resolve the dense clumps embedded in GMCs, providing crucial information about massive star formation.

\begin{figure*}[htbp]
  \begin{center}
  \includegraphics[width=17cm]{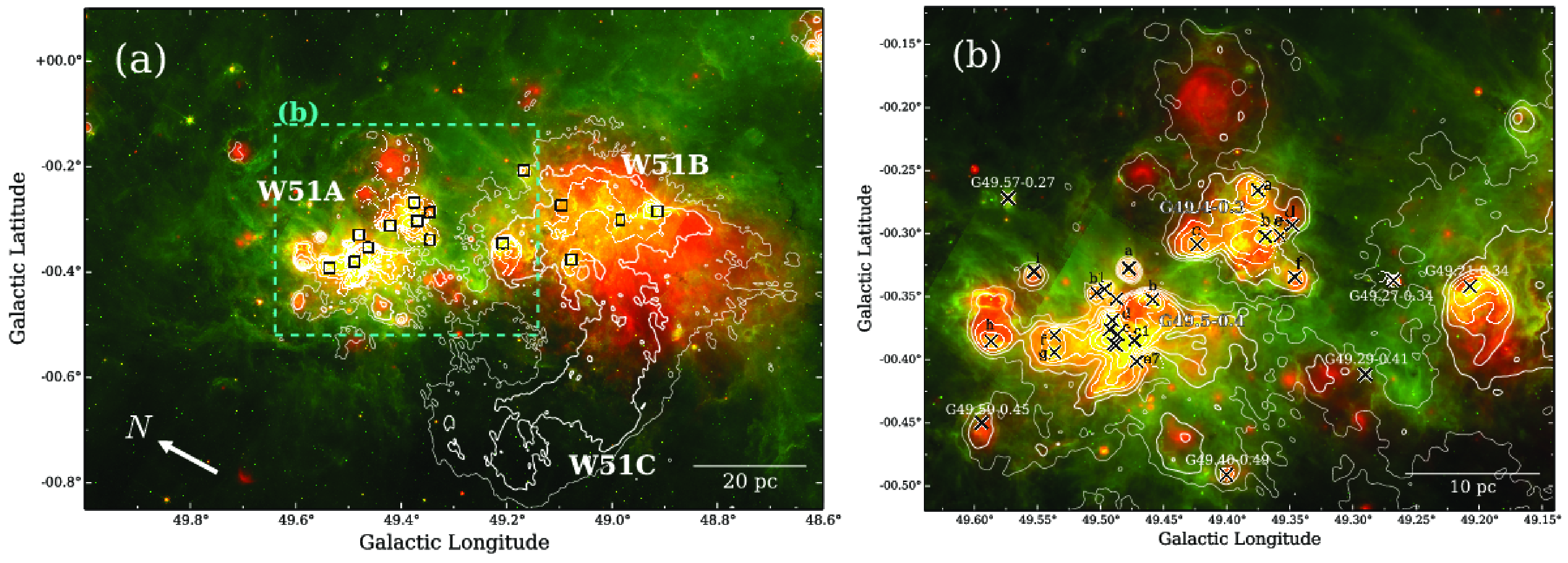}
  \end{center}
  \caption{(a) A composite color image of the {\it Spitzer}/MIPSGAL 24 $\mu$m (red) and {\it Spitzer}/GLIMPSE 8$\mu$m (green) emissions toward W51. The white contours indicate \textcolor{black}{the THOR 21\,cm radio continuum emission combined with the VGPS data (\cite{2016A&A...595A..32B, 2006AJ....132.1158S}), and are plotted from 0.03 (dashed lines) to 3.0\,Jy\,str$^{-1}$ with logarithm step. The angular resolution of the THOR data combined with the VGPS is 25$''$. } Square symbols represent the compact radio continuum sources identified by \citet{1997ApJS..108..489K}. \textcolor{black}{FUGIN data cover the entire region of this map. } (b) A close-up view of W51A. The corresponding region is indicated by the box with dashed blue lines in (a). Crosses represent the H{\sc ii} regions listed in \citet{1994ApJS...91..713M} (see also Table\,\ref{tab:sources}).}\label{fig:intro_fig}
\end{figure*}

W51 is one of the most active massive star-forming regions in the Milky Way (MW).
It was discovered by the observations of the thermal radio continuum emission at 21\,cm \citep{1958BAN....14..215W}.
The distance to W51 was measured as $5.4\pm0.3$\,kpc by \citet{2010ApJ...720.1055S} based on the observations of trigonometric parallax, and the total far-infrared luminosity of W51 is as high as $\sim 8 \times 10^6\, L_{\odot}$ at 5.4\,kpc \citep{1984ApJ...286..573R}. 
As shown in Figure\,\ref{fig:intro_fig}, which shows a two-color composite image of W51 with the {\it Spitzer} 8 and 24\,$\mu$m data overlaid with the contour map of the 21\,cm emission, W51 consists of a number of H{\sc ii} regions for a large area of $\sim1^\circ$, which corresponds to $\sim100$\,pc at 5.4\,kpc, and these H{\sc ii} regions are separated into two major groups, called W51A and W51B, in the northeastern and southwestern parts of W51, respectively \citep{1975LNP....42..443B}. 
\textcolor{black}{The 21\,cm radio continuum emission data was taken from THOR (The H{\sc i}, OH, Recombination line survey of the Milky Way) archive, which is combined with the VGPS data (\cite{2016A&A...595A..32B, 2006AJ....132.1158S}). }
The total stellar masses included in W51A and W51B were measured as $1.8 \times 10^4\ M_{\odot}$ and $1.4 \times 10^4\ M_{\odot}$, respectively \citep{2000ApJ...543..799O, 2007JKAS...40...17K}.
A supernova remnant W51C is located in the southeast, which can be traced in the non-thermal radio continuum emission (\cite{1997ApJ...475..194K, 1997ApJ...485..263K}). 
In this study, we focus on W51A, as it is a young massive star-forming region with ages of a few 0.1--1\,Myrs (e.g., \cite{2004MNRAS.353.1025K, 2015A&A...573A.106G}), providing a unique opportunity to investigate \textcolor{black}{mechanisms yielding one of the highest star-forming rate in our Galaxy (\cite{2000ApJ...543..799O})}.

W51A harbors two bright H{\sc ii} region complexes, \textcolor{black}{GAL 049.5-00.4 and GAL 049.4-00.3 (hereafter called ``G49.5-0.4" and ``G49.4-0.3")}, which each include several H{\sc ii} regions within $\sim$10\,pc, as shown in the 21\,cm continuum emissions in Figure\,\ref{fig:intro_fig}(b). 
In G49.5-0.4, sixteen H{\sc ii} regions, named G49.5-0.4a--i, were identified \citep{1994ApJS...91..713M}. 
The $J$, $H$, $K'$, and Br$\gamma$ photometric observations by \citet{2000ApJ...543..799O} identified many O-type and early B-type stars toward these H{\sc ii} regions, as summarized in Table\,\ref{tab:sources}.
The sizes of these H{\sc ii} regions range from $\sim$0.1\,pc to $\sim$2\,pc, and these sizes were used to estimate the ages of these H{\sc ii} regions to be 0.1--2.6\,Myr \citep{2000ApJ...543..799O}. 
The two outstanding radio sources G49.5-0.4e and G49.5-0.4d are also known as W51\,IRS\,1 and IRS\,2 (\cite{1974ApJ...187..473W}), which each harbor 4--5 O-type stars and 2--4 early B-type stars, forming bright stellar clusters at the center of G49.5-0.4.
The other sources G49.5-0.4a, b, c1, f, h, and i also have multiple O-type/early B-type stars.
In addition to these H{\sc ii} regions, many massive young stellar objects (MYSOs) have been identified throughout G49.5-0.4 (\cite{2009ApJ...706...83K, 2017ApJ...839..108S}), indicating that massive star formation still continues in this region.

On the other hand, G49.4-0.3 consists of six H{\sc ii} regions G49.4-0.3 a--f with sizes of $\sim$0.7--2\,pc \citep{1994ApJS...91..713M}. 
Although there is no photometric identification of the exciting stars of these H{\sc ii} regions, their classifications were estimated to be O4--B0 as listed in Table\,\ref{tab:sources} from the measurements of ionization photons with the 21\,cm radio continuum data (\cite{1997ApJS..108..489K}).
The typical age of the H{\sc ii} regions was measured as $\sim$0.2\,Myrs (see Section\,5.1).
Other than G49.5-0.4 and G49.4-0.3, there are several discrete H{\sc ii} regions within $\sim$20\,pc from G49.5-0.4 or G49.4-0.3, and the classifications of the exciting stars in these sources were measured as O4--B0 from the 21\,cm data (see a summary in Table\,\ref{tab:sources}).

\begin{table}[h]
   \tbl{List of the H{\sc ii} regions and massive stars in W51A}
   {%
    \begin{tabular}{c|c|c|c|c|c}
\hline \hline
Name & \textcolor{black}{Radius} & log $N_{\rm i}$ & Classification & \textcolor{black}{Age} & Reference \\
 & (pc) & (photons/sec) &  & (Myr) & \\ 
(1)  & (2) & (3) & (4) & (5) & (6) \\ 
 \hline \hline
G49.21-0.34 & 4.9 & 49.8  & [O4] & 2.2*  & [1]  \\
G49.27-0.34 & 0.9 & 45.6  & [B0] & 1.3*  & [1]  \\
G49.29-0.41 & -- & 47.3  & [B0] & --  & [1]  \\
G49.4-0.3 a & 1.6 & 49.0  & [O6] & 0.5*  & [1, 2] \\
... b & 0.4 & 49.7  & [O5] & 0.1*  & [1, 2] \\
... c & 1.6 & 49.3  & [O5.5] &0.4*  & [1, 2]  \\
... d & 1.6 & 48.6  & [O7] & 0.6* & [1, 2] \\
... e & 0.3 & 47.7  & [O9.5] & 0.1* & [1, 2]  \\
... f & 1.6 & 48.2  & [O8.5] & 0.8* & [1]  \\
G49.40-0.49 & 1.3 & 48.2  & [O8.5] & 0.5* & [1]  \\
G49.5-0.4 a & 1.6 & 49.0  & O5+B1  & 0.7  & [1, 2, 3]\\
... b & 1.6 & 49.6  & O4+O8+B0 & 2.2  & [1, 2, 3]\\
... b1 & 0.7 & 48.2  & O9 & 0.8   & [1, 3]\\
... b2 & 0.5 & 47.3  & B1 & 0.2   & [1, 3]\\
... b3 & 0.5 & 48.1  & [O9] & -   & [1] \\
... c1 & -- & 49.2  & O5+O6+B0 & 0.4   & [1, 3] \\
... d (IRS 2) & 1.1 & 49.7  & O4+O6$\times$4+B0+B1$\times$3& 0.1  & [1, 3] \\
... e (IRS 1) & 0.4 & 50.4  &  O4+O5+O6+O8+B0$\times$2 & -  & [1, 2, 3] \\
... e1 & -- & 47.9  & [O9.5] & -   & [1] \\
... e2 & -- & $>$47.6 & [$>$B0] & -   & [1] \\
... e6 & -- & 47.3  & [B0] & -   & [1] \\
... e7 & -- & 47.2  & B0 & 0.3   & [1, 3]\\
... f & 1.9 (f+g) & 49.1 & O7+B0+B1$\times$2 & 1.8   & [1, 2, 3]\\
... g & 1.9 (f+g) & 48.9  & [O5$\times$2+B1] & 1.5 & [2] \\
... h & 3.2 & 48.9  & O5+B0$\times$4+B1 & 2.6  & [1, 2, 3]\\
... i & 1.6 & 48.4  & O8+B1 & 1.5 & [1, 3] \\
G49.57-0.27 & 1.2 & 47.5  & [B0] & 0.7* & [1] \\
G49.59-0.45 & 3.3 & 48.7  & [O7] & 2.1* & [1] \\
 \hline
    \end{tabular}} \label{tab:sources}
\begin{tabnote}
(1) Name of the H{\sc ii} region. 
(2) \textcolor{black}{Radius of the H{\sc ii} region taken from the web site of the Wide-field Infrared Survey Explorer (WISE) catalog of Galactic H{\sc ii} regions (\cite{2014ApJS..212....1A, 2017ApJ...846...64M}).} 
(3) Ionizing photon flux estimated by \citet{1994ApJS...91..713M}.
(4) Classification of the exciting stars in the H{\sc ii} region estimated by \citet{1994ApJS...91..713M} and \citet{2000ApJ...543..799O}. Those derived by measuring ionizing photons from radio continuum image is shown with brackets, while those identified in the near-infrared photometric observations are presented without brackets.
(5) Expansion age of the H{\sc ii} region estimated by \citet{2000ApJ...543..799O} or in this study. Those with asterisk were measured in this study (see Section\ref{sec:age}).
(6) References. 
[1] \citet{1994ApJS...91..713M}, 
[2] \citet{1997ApJS..108..489K}, 
[3] \citet{2000ApJ...543..799O}.
 
\end{tabnote}
\end{table}

Molecular gas in W51 shows an extended distribution for $\sim 100$\,pc\,$\times$\,100\,pc, with a total molecular mass measured as $\sim 7.1 \times 10^5\ M_{\odot}$ at 5.4\,kpc \citep{1998AJ....116.1856C}.
The CO emissions in W51A have multiple velocity components in a radial velocity \textcolor{black}{(line-of-sight velocity)} range between $\sim$50--70\,km\,s$^{-1}$.
Based on the large-scale $^{12}$CO and $^{13}$CO ($J$=1--0) observations covering the entirety of the W51 region at an angular resolution of 46$''$, \citet{1998AJ....116.1856C} decomposed the velocity structures of molecular gas by performing fits to the CO spectra with multiple Gaussian functions.
Subsequently, \citet{2001PASJ...53..793O} performed $^{13}$CO ($J$=1--0) observations at a high angular resolution of $15''$ toward a $\sim 15' \times 15'$ area centered on G49.5-0.4. By analyzing the position-velocity diagrams, the authors identified four velocity components around radial velocities of 50, 56, 60, and 68\,km\,s$^{-1}$.

The 68\,km\,s$^{-1}$ cloud corresponds to High Velocity Stream (HVS), which is a filamentary molecular cloud stretched nearly parallel to the Galactic plane, overlapping W51A and W51B along the line of sight. 
The length and width of HVS were measured as $\sim 100$\,pc and $\sim 10$\,pc, respectively \citep{1998AJ....116.1856C, 2010ApJS..190...58K, 2012MNRAS.424.1658P}.
\citet{1970IAUS...38..397B} and \citet{1999ApJ...518..760K} discussed the large velocity of HVS as being attributed to the streaming motion of gas down to the Sagittarius spiral arm driven by the spiral density wave.
For the velocity components other than HVS, \citet{1998AJ....116.1856C} discussed that these represent kinematic structures within a single molecular cloud, the W51 cloud, with a total molecular mass of $\sim 6.0 \times 10^5\,M_{\odot}$, referred to as the W51 cloud, whereas \citet{2001PASJ...53..793O} postulated that these are discrete molecular clouds located at the same distance.

It has been actively debated that the massive star formation in W51A was triggered by collisions between molecular clouds having different radial velocities over $\sim$20\,km\,s$^{-1}$ \citep{1979A&A....75..365P, 1985A&A...145..369A, 1998AJ....116.1856C, 1999ApJ...518..760K, 2001PASJ...53..793O}. 
\citet{1998AJ....116.1856C} proposed a collision between the W51 cloud and HVS. 
The authors revealed that the CO emissions around 60\,km\,s$^{-1}$ in the W51 cloud truncate at the location of HVS, and discussed these two velocity components as being physically related objects at a common distance, suggesting a collision between these two clouds.
\citet{2010ApJS..190...58K} reached the same conclusion, based on the $^{12}$CO and $^{13}$CO ($J$=2--1) observations at 36$''$ resolution, which covered a $1\fdg25 \times 1\fdg00$ area of W51.
\citet{2001PASJ...53..793O} argued that a ``pileup'' scenario of the four discrete molecular clouds resulted in a burst of massive star formation in G49.5-0.4 in W51A. 

Recently, supersonic collision between molecular clouds has been discussed as a plausible mechanism of massive star formation.
These observational studies of cloud-cloud collisions (CCCs) include the super star clusters and the H{\sc ii} regions in the MW and young O stars in the Large Magellanic Cloud \citep{2014ApJ...780...36F, 2015ApJ...807L...4F, 2017arXiv170605768F,  2018ApJ...859..166F, 2009ApJ...696L.115F, 2017arXiv170605871H, 2017arXiv170607964K, 2017arXiv170606956N, 2017arXiv170606002N, 2010ApJ...709..975O, 2017arXiv170605659O, 2017arXiv170605652O, 2017arXiv170605763S, 2013ApJ...768...72S, 2011ApJ...738...46T, 2015ApJ...806....7T, 2017ApJ...835..142T, 2017arXiv170607164T, 2015PASJ...67..109T, 2017arXiv170608656T, 2017arXiv170605664T}, where the super star clusters include 10--20 O stars, while the others include a single young O star.
Formation of the massive clumps, which may form massive stars, in the collisional-compressed layer was discussed in depth in the magneto-hydrodynamical (MHD) simulations by \citet{2013ApJ...774L..31I} and \citet{2017arXiv170702035I}. 
\citet{2017arXiv170807952K} formulated the time evolution equation of GMC mass function including CCC, indicating that CCC-driven star formation is mostly driven by massive GMCs having masses $>10^{5.5}$\,$M_\odot$, which may account for a few 10\% of the total star formation in the MW and nearby galaxies.
Comparisons between the observations and numerical calculations have indicated two important observational signatures of CCCs, i.e., ``broad bridge feature'' in position-velocity diagrams and ``complementary distribution'' on the sky between two molecular clouds with different velocities, which provide useful diagnostics to investigate CCCs with molecular line observations (\cite{1992PASJ...44..203H, 2010MNRAS.405.1431A, 2014ApJ...792...63T, 2017ApJ...835..142T, 2018ApJ...859..166F}).

\textcolor{black}{\subsection{Observational signatures of CCC}}\label{sec:obsCCC}

Based on comparisons between observations and simulations, \citet{2018ApJ...859..166F} and \citet{2017ApJ...835..142T} discussed two possible observational signatures of CCC, i.e., ``broad bridge feature'' in position-velocity diagrams and ``complementary distribution'' on the sky between two clouds with different velocities, where the authors assumed a collision between two dissimilar clouds based on the basic CCC \textcolor{black}{scenarios} studied by \citet{1992PASJ...44..203H}, followed by \citet{2010MNRAS.405.1431A, 2014ApJ...792...63T, 2015MNRAS.454.1634H, 2015MNRAS.450...10H, 2017arXiv170608656T}.
A broad bridge feature is relatively weak CO emissions at intermediate velocities between two colliding clouds that are separated in velocity.
When a smaller cloud drives into a larger cloud, a dense compressed layer at the collisional interface is formed, resulting in a thin turbulent layer between the larger cloud and the compressed layer. 
If one observes a snapshot of this collision with a viewing angle parallel to the colliding axis, two velocity peaks separated by intermediate-velocity emission with lower intensity can be seen in the position-velocity diagrams. 
The turbulent gas which creates the broad bridge feature can be replenished as long as the collision continues.
Several observational studies reported detections of broad bridge features in the CCC regions (e.g., \cite{2014ApJ...780...36F, 2016ApJ...820...26F, 2017PASJ...69L...5F, 2009ApJ...696L.115F, 2010ApJ...709..975O, 2015ApJ...806....7T, 2017ApJ...835..142T}).

When two clouds collide, one caves the other owing to the momentum conservation (\cite{2015MNRAS.450...10H}).
If the collision takes place head-on between two dissimilar clouds, a cavity will be formed on the larger cloud through this process, and the larger cloud can be seen as a ring-like structure on the sky, unless the observer viewing angle is perfectly perpendicular to the colliding axis.
As the size of the cavity corresponds to that of the smaller cloud, the observer with a viewing angle parallel to the colliding axis sees a complementary distribution between the smaller cloud and the ring-like structure of the larger cloud.
If the collision is an offset collision, not a head-on collision, the basic process is not changed, and two clouds with different velocities can be observed close to each other. These two clouds may share the boundaries of the clouds on the sky, showing a complementary distribution.
\citet{2017arXiv170605768F} and \citet{2018ApJ...859..166F} pointed out that if the observer viewing angle has an inclination relative to the colliding axis, the complementary distribution has a spatial offset depending on the travel distance of the collision or the depth of the cavity.
 
In the well-resolved CCC regions, a combination of the two signatures of CCC, broad bridge feature and complementary distribution, may be observed as a ``V-shape'' gas distribution in the $p$-$v$ diagram (e.g., \cite{2018ApJ...859..166F, 2017arXiv170605652O, 2017arXiv170605871H, 2017arXiv170607164T}).
Analyses of the synthetic CO data by \citet{2018ApJ...859..166F} indicate that, if the observer viewing angle is inclined relative to the colliding axis, the V-shape distribution becomes skewed.


\textcolor{black}{\subsection{Paper overview}}\label{sec:overview}

Following the recent improvement of our knowledge on CCC as a trigger of massive star formation, in this study we present an analysis of the new $^{12}$CO, $^{13}$CO, and C$^{18}$O ($J$=1--0) data covering the entirety of W51A in order to test the CCC \textcolor{black}{scenarios} as the mechanism of the active massive star formation in W51A.
The CO data was obtained using the Nobeyama 45-m telescope at 20$''$ resolution, which corresponds to $\sim$0.5\,pc at 5.4\,kpc, as a part of the Galactic plane survey legacy project FUGIN (FOREST Unbiased Galactic plane Imaging survey with the Nobeyama 45-m telescope) (\cite{2016SPIE.9914E..1ZM, 2017PASJ...69...78U}).
The advantages of our new CO ($J$=1--0) data can be summarized as follows:
\begin{enumerate}
\item We covered a large area of $1\fdg 4 \times 1\fdg 0$ including W51A at a comparable spatial resolution with that in the $^{13}$CO ($J$=1--0) observations by \citet{2001PASJ...53..793O}, which covered a $\sim 15' \times 15'$ area of G49.5-0.4. 
\item Our data includes the C$^{18}$O ($J$=1--0) emission, which allows us to diagnose the signatures of CCCs in the molecular clouds in W51A. Note that the C$^{18}$O ($J$=1--0) emission has not been studied for a large area of W51A at such a high angular resolution. \citet{2012MNRAS.424.1658P} performed a large-scale $^{12}$CO, $^{13}$CO, and C$^{18}$O ($J$=3--2) observations with the James Clerk Maxwell Telescope (JCMT) toward W51A and W51B, providing a comprehensive catalog of the dense gas in the molecular clouds in W51A. However, the authors did not focus on the spatial and velocity distributions of gas with the aim of investigating interactions among different velocity components. 
\item Our CO data has a comparable spatial resolution with the JCMT archival CO ($J$=3--2) data \citep{2012MNRAS.424.1658P}, allowing us to investigate the excitation conditions of gas, and to probe interaction between molecular gas and H{\sc ii} regions.
\end{enumerate}
 In Section\,2 we describe the CO dataset used in this study, and in Section\,3 we present the main results of the analyses on the CO dataset and comparisons with the other wavelengths. 
In Section\,4 we discuss the results, and present a summary in Section\,5.

\section{Dataset}
The observations of W51A were carried out as a part of FUGIN project (\cite{2017PASJ...69...78U}) \textcolor{black}{with Nobeyama Radio Observatory (NRO) 45-m telescope}.
Details of the observations, calibration, and data reduction are summarized in \citet{2017PASJ...69...78U}, and parameters of the observations and output data are listed in Table \ref{tab:obs}. 
In W51A, we covered a  $l = 50\fdg 0$--$48\fdg 6$, $b = -0\fdg 9$--$+0\fdg 1$ ($1\fdg 4 \times 1\fdg 0$) in the $^{12}$CO ($J$=1--0), $^{13}$CO ($J$=1--0), and C$^{18}$O ($J$=1--0) emissions.
\textcolor{black}{A beam size of NRO 45-m telescope is $\sim 15''$ at 115\,GHz and an effective angular resolution of this mapping is $\sim 20''$. }
The SAM45 (Spectral Analysis Machine for the 45-m telescope) spectrometer (\cite{2011KUNO}) was used at a frequency resolution of 244.14\,kHz, and the effective velocity resolution was 1.3\,km\,s$^{-1}$ at 115\,GHz.
The typical system noise temperatures including atmosphere were $\sim$150\,K and $\sim$250\,K at 110\,GHz and 115\,GHz, respectively.  
The output cube data has spatial grids of $8.5'' \times 8.5''$ and velocity grid of 0.65\,km\,s$^{-1}$ \textcolor{black}{for the $^{12}$CO ($J$=1--0), $^{13}$CO ($J$=1--0), and C$^{18}$O ($J$=1--0) emissions}.
The final root-mean-square (r.m.s) noise temperature $T_{\rm{rms}}$ in $T_{\rm{mb}}$ scale are 1.5\,K, 0.7\,K, and 0.7\,K \textcolor{black}{per velocity channel} for $^{12}$CO ($J$=1--0), $^{13}$CO ($J$=1--0), and C$^{18}$O ($J$=1--0), respectively. 

The $^{12}$CO ($J$=3--2), $^{13}$CO ($J$=3--2), and C$^{18}$O ($J$=3--2) data were obtained by \citet{2012MNRAS.424.1658P} with the Heterodyne Array Receiver Programme (HARP) receiver and the back-end digital autocorrelator spectrometer Auto-Correlation Spectral Imaging System (ACSIS) on the JCMT.
The observations covered a $1\fdg 4 \times 1\fdg 0$ area including W51A and W51B.
The data has an angular resolution of 14$''$ and a velocity resolution of 0.5 km s$^{-1}$.

\textcolor{black}{In figures in this paper, to improve a signal-to-noise ratio and compare the FUGIN data with the JCMT data on the same angular resolution, we convolved the dataset with Gaussian of FWHM 22.4$''$ and 26.5$''$ for the FUGIN data and the JCMT data, respectively (giving the smoothed angular resolutions of $\sqrt{20.0^2+22.4^2}\approx30.''0$ and $\sqrt{14.0^2+26.5^2}\approx30.''0$). 
We also convolved the dataset for the velocity axis to be a resolution of 1.3\,km\,s$^{-1}$ using the same method. }
 
\begin{table*}[h]
  \tbl{Summary of the CO ($J$=1--0) dataset }{%
  \begin{tabular}{l|l}
      \hline
       Observation Date & March -- May 2014 and April -- May 2015 \\
       Observed Area &  $l = 50\fdg 0$--$48\fdg 6$, $b = -0\fdg 9$--$+0\fdg 1$ ($1\fdg 4 \times 1\fdg 0$)\\
       Telescope & NRO 45-m telescope \\
       Receiver & FOREST \\
       Observation Mode & On-The-Fly \\
       Emission Lines & $^{12}$CO ($J$=1--0), $^{13}$CO ($J$=1--0), and C$^{18}$O ($J$=1--0) \\
       Angular and Velocity Resolution & $\sim20''$ ($\sim 0.5$\,pc for distance 5.4\,kpc) and $\sim1.3$\,km\,s$^{-1}$\\
       Angular and Velocity Grid of the final cube data & $8.5''$ and $0.65$ km $\rm{s^{-1}}$\\
       $T_{\rm{rms}}$ & $\sim\,1.5$\,K for $^{12}$CO ($J$=1--0), $\sim\,0.7$\,K for $^{13}$CO ($J$=1--0), and $\sim\,0.7$\,K for C$^{18}$O ($J$=1--0)\\
       \hline
    \end{tabular}}\label{tab:obs}
\end{table*}

\clearpage

\section{Results}\label{sec:res}

\subsection{Large-scale gas distribution}
\subsubsection{CO ($J$=1--0) distribution}
\begin{figure}[h]
 \begin{center}
  \includegraphics[width=17cm]{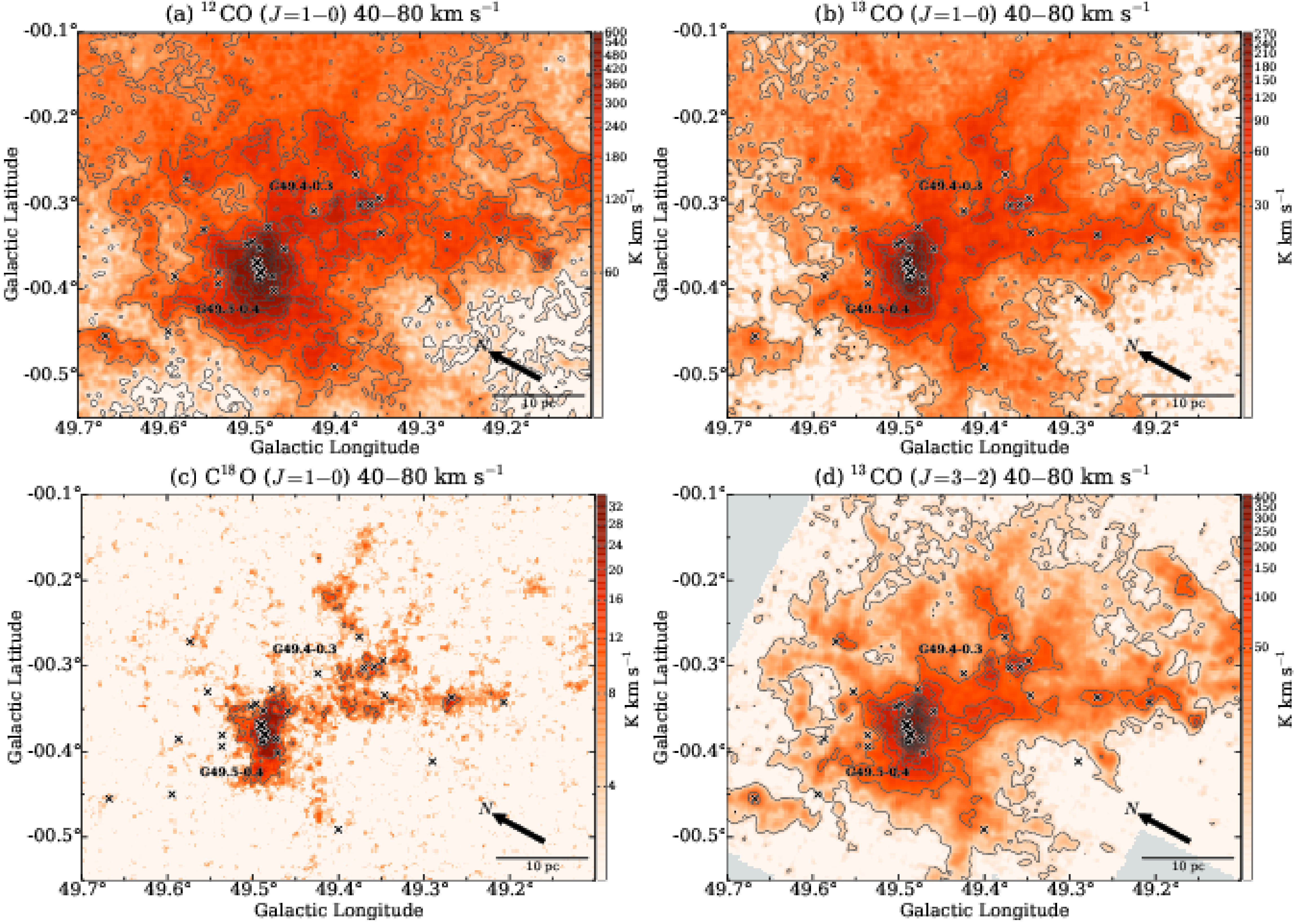}
 \end{center}
 \caption{\textcolor{black}{Integrated intensity map of (a) $^{12}$CO ($J$=1--0), (b) $^{13}$CO ($J$=1--0), (c) C$^{18}$O ($J$=1--0), and (d) $^{13}$CO ($J$=3--2) integrated over 40--80\,km\,s$^{-1}$ toward W51A. These maps are colored in logarithmic scale. The contours are plotted with (a) $9\,\sigma$ (60\,K\,km\,s$^{-1}$) intervals starting from the $3\,\sigma$ (20\,K\,km\,s$^{-1}$) level, (b) $9\,\sigma$ (30\,K\,km\,s$^{-1}$) intervals starting from the $3\,\sigma$ (10K\,km\,s$^{-1}$) level, (c) $3\,\sigma$ (10\,K\,km\,s$^{-1}$) intervals starting from the $3\,\sigma$ (10\,K\,km\,s$^{-1}$) level, and (d) $2\,\sigma$ (30\,K\,km\,s$^{-1}$) intervals starting from the $3\,\sigma$ (50\,K\,km\,s$^{-1}$) level, respectively. The gray pixel in (d) indicate the area not covered by JCMT \citep{2012MNRAS.424.1658P}.}}\label{fig:integ_all}
\end{figure}

\textcolor{black}{Figures\,\ref{fig:integ_all}(a)--(d) show the integrated intensity maps of the $^{12}$CO ($J$=1--0), $^{13}$CO ($J$=1--0), C$^{18}$O ($J$=1--0), and $^{13}$CO ($J$=3--2) emissions integrated over 40 -- 80\,km\,s$^{-1}$, which covers the W51A region shown in Figure\,\ref{fig:intro_fig}(b). 
In Figure\,\ref{fig:integ_all}(a) the $^{12}$CO ($J$=1--0) emissions show extended gas distributions \textcolor{black}{over the entire map}, while the $^{13}$CO ($J$=1--0) emissions presented in Figure\,\ref{fig:integ_all}(b) show \textcolor{black}{somewhat clumpy} structures.
In Figure\,\ref{fig:integ_all}(c) C$^{18}$O ($J$=1--0) emissions show \textcolor{black}{more} clumpy structures \textcolor{black}{than the $^{13}$CO ($J$=1--0) emissions}.
They are strongly detected in G49.5-0.4, indicating the presence of high density gas in in this region. 
The distribution of $^{13}$CO ($J$=3--2) emissions (JCMT data obtained by \citet{2012MNRAS.424.1658P}) presented in Figure\,\ref{fig:integ_all}(d) is resembles that of $^{13}$CO ($J$=1--0) emissions, but the intensity is relatively low in the northwestern part of the map. }

\begin{figure}[h]
 \begin{center}
  \includegraphics[width=17cm]{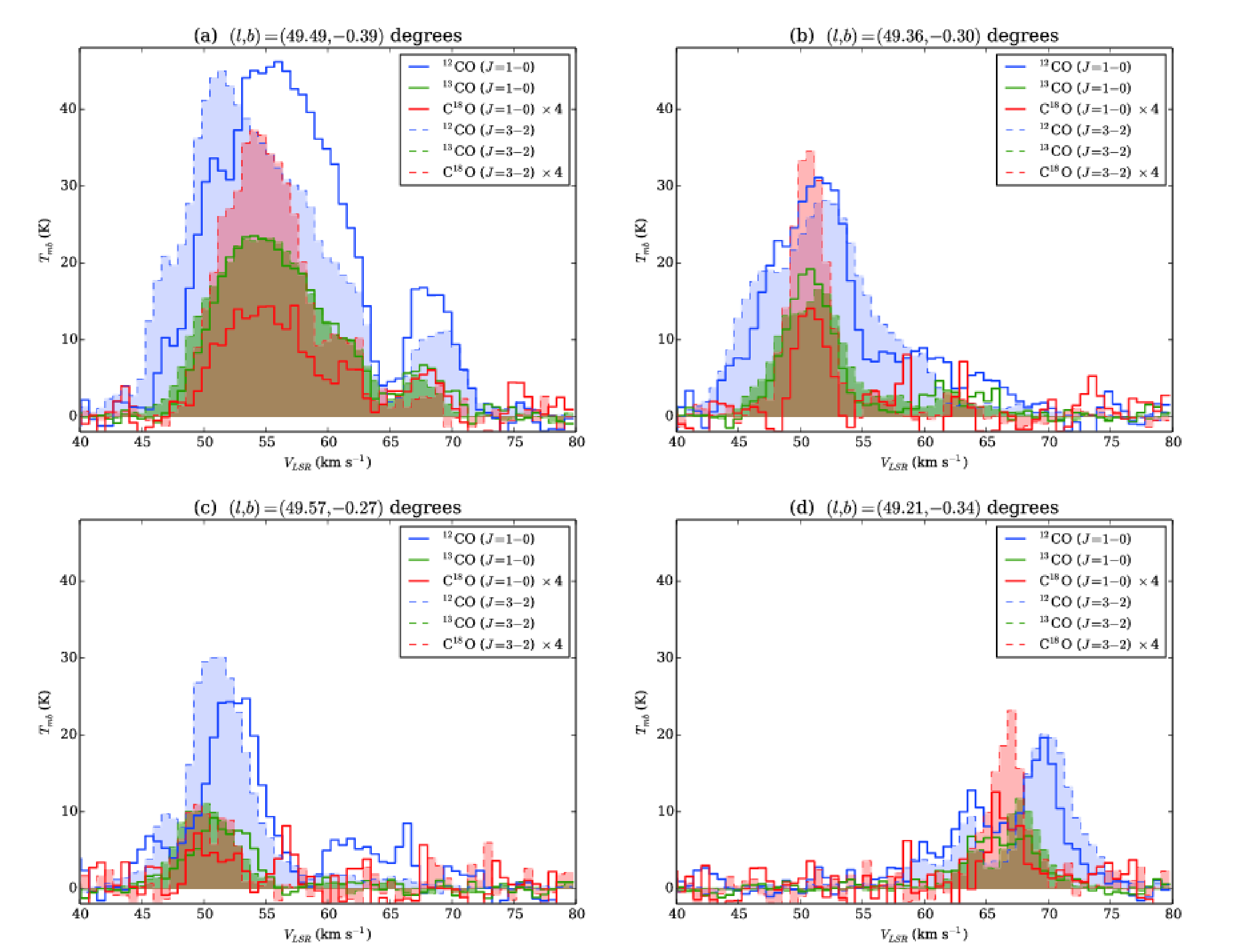}
 \end{center}
 \caption{\textcolor{black}{The CO spectra at (a) $(l,b)=(49\fdg 49, -0\fdg 39)$, (b) $(l,b)=(49\fdg 36, -0\fdg 30)$, (c) $(l,b)=(49\fdg 57, 0\fdg 27)$, and (d) $(l,b)=(49\fdg 21, -0\fdg 34)$. Blue, green, and red lines indicate the profiles of the $^{12}$CO, $^{13}$CO, and C$^{18}$O emission lines, respectively. \textcolor{black}{The intensities of the C$^{18}$O emission lines were multiplied by 4.} Solid lines and broken lines \textcolor{black}{(filled)} indicate the profiles of the $J$=1--0 and $J$=3--2 emission lines, respectively. }\textcolor{black}{The spatial grid size and resolution are $8.''5$ ($=0\fdg 0023$) and $\sim30''$ ($=0\fdg 0083$), respectively. }}\label{fig:spe}
\end{figure}

\textcolor{black}{Figures\,\ref{fig:spe}(a)--(d) show the spectra of the $^{12}$CO ($J$=1--0), $^{13}$CO ($J$=1--0), C$^{18}$O ($J$=1--0) (FUGIN data), $^{12}$CO ($J$=3--2), $^{13}$CO ($J$=3--2), and C$^{18}$O ($J$=3--2) (JCMT data) toward four representative positions. 
Figure\,\ref{fig:spe}(a) shows the spectra at the $^{13}$CO ($J$=1--0) peak \textcolor{black}{positions} in G49.5-0.4. 
CO emissions are detected in the velocity range of 45--65\,km\,s$^{-1}$ and $\sim$\,68\,km\,s$^{-1}$ with complicated spectral profiles. 
Figure\,\ref{fig:spe}(b) shows the spectra at the $^{13}$CO ($J$=1--0) peak \textcolor{black}{positions} in G49.4-0.3. 
The profiles of the \textcolor{black}{spectra} have a peak at $\sim$51\,km\,s$^{-1}$, while weak CO emissions are detected around $\sim$60\,km\,s$^{-1}$. 
The CO emissions at $\sim$68\,km\,s$^{-1}$ seen toward G49.4-0.3 are barely detected. 
Figure\,\ref{fig:spe}(c) and Figure\,\ref{fig:spe}(d) shows the spectra toward the H{\sc ii} region G49.57-0.27 and G49.34-0.21, respectively. 
CO emissions are detected in several velocity ranges.
In addition, in Figure\,\ref{fig:spe}(d), we can see that the $^{12}$CO ($J$=1--0) and $^{12}$CO ($J$=3--2) \textcolor{black}{spectra} show self-absorption features at $\sim$66--68\,km\,s$^{-1}$ indicated by the C$^{18}$O emissions at the velocity. }

\begin{figure}[h]
 \begin{center}
  \includegraphics[width=17cm]{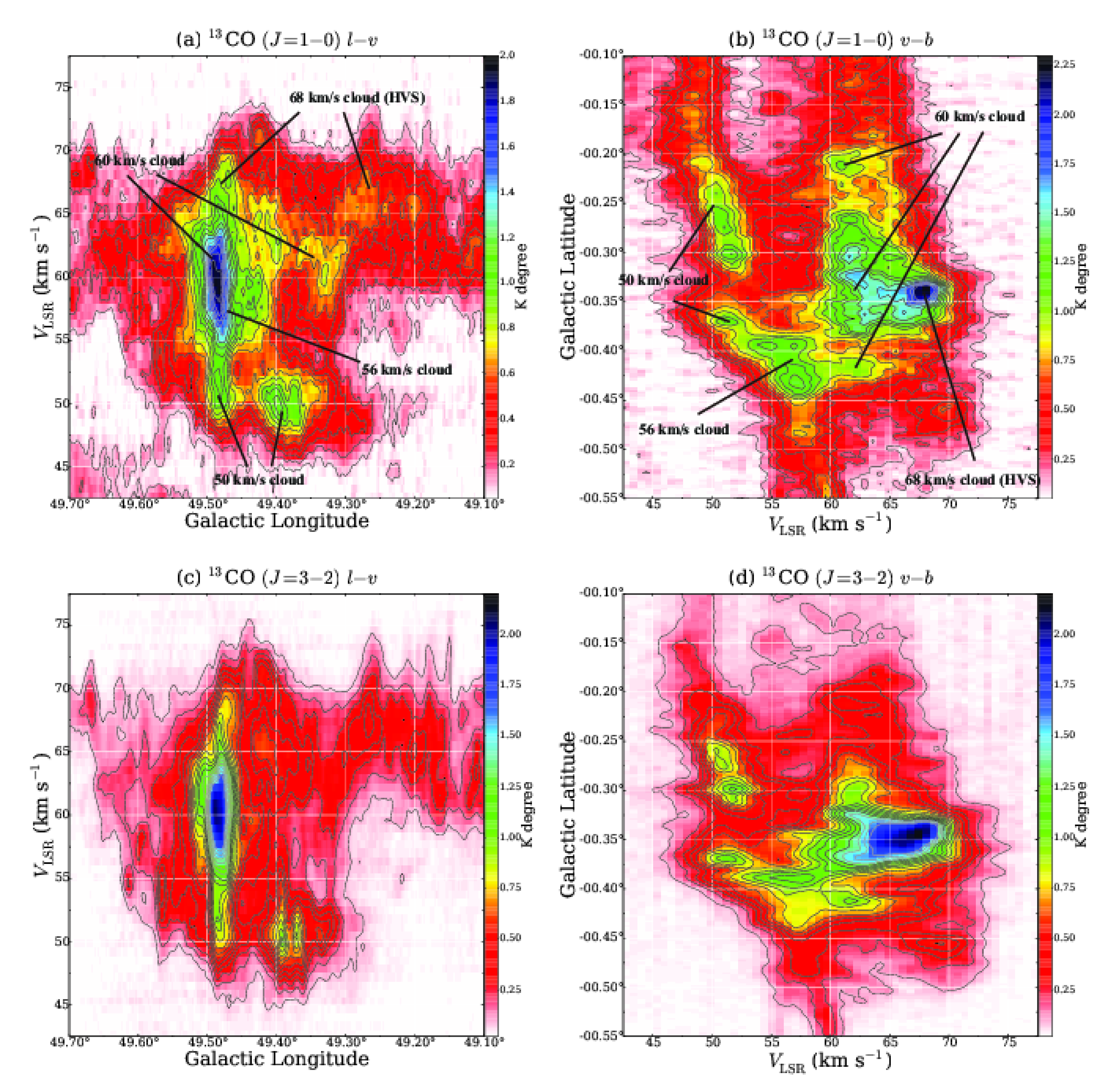}
 \end{center}
 \caption{\textcolor{black}{(a) The Galactic Latitude--Velocity ($l-v$) diagram of the $^{13}$CO ($J$=1--0) emissions integrated over $b=-0\fdg 55$ -- $-0\fdg 10$. The contours are plotted with $5\,\sigma$ (0.11\,K\,degrees) intervals starting from the $5\,\sigma$ (0.11\,K\,degrees) level. (b) The Velocity--Galactic Latitude ($v-b$) diagram of the $^{13}$CO ($J$=1--0) emissions integrated over $l=49\fdg 70$ -- $49\fdg 10$. The contours are plotted with $5\,\sigma$ (0.13\,K\,degrees) intervals starting from the $5\,\sigma$ (0.13\,K\,degrees) level. (c) Same as (a), but for the $^{13}$CO ($J$=3--2) emissions. The contours are plotted at every $5\,\sigma$ (0.09 \,K\,degrees) from 0.09 \,K\,degrees ($\sim 5\sigma$). (d) Same as (b), but for the $^{13}$CO ($J$=3--2) emissions. The contours are plotted with $5\,\sigma$ (0.11\,K\,degrees) intervals starting from the $5\,\sigma$ (0.11\,K\,degrees) level. }
 }\label{fig:lvs}
\end{figure}

\textcolor{black}{Figures\,\ref{fig:lvs}(a)--(d) show the $l$-$v$ diagram (integrated along the Galactic latitude over $b = -0\fdg55$ -- $-0\fdg10$) and the $v$-$b$ diagram (integrated along the Galactic longitude over $l = 49\fdg70$ -- $49\fdg10$) of the $^{13}$CO ($J$=1--0) emissions and the $^{13}$CO ($J$=3--2) emissions.}
Four discrete velocity components identified in the previous studies \citep{2001PASJ...53..793O} can be seen in the spectra, $l$-$v$ diagram, and  $v$-$b$ diagram, and these clouds are connected with each other by the intermediate velocity emissions.
Following the nomenclature by \citet{2001PASJ...53..793O}, we hereafter refer to the velocity components around 50, 56, 60, and 68\,km\,s$^{-1}$ as ``the 50\,km\,s$^{-1}$ cloud'', ``the 56\,km\,s$^{-1}$ cloud'', ``the 60\,km\,s$^{-1}$ cloud'', and ``the 68\,km\,s$^{-1}$ cloud (HVS)'', respectively. 

\textcolor{black}{Besides these four clouds, we can see a molecular cloud nearby G49.5-0.4 at higher velocity range (71.6 -- 74.8\,km\,s$^{-1}$) in the velocity channel maps of the $^{12}$CO ($J$=1--0) and $^{13}$CO ($J$=1--0) emission (Figure\,\ref{fig:chmap} in Appendix\,\ref{app:chmap}). 
It extends perpendicular to the elongation of the 68\,km\,s$^{-1}$ cloud (HVS) down to $b = \sim -0\fdg5$. 
Its elongation is perhaps related to feedback of the massive stars in G49.5-0.4. 
In this paper, we disregard the molecular clouds in this velocity range because their intensity of CO emissions is significantly lower than the other four velocity components. }

\begin{figure*}[htbp]
  \begin{center}
         \includegraphics[width=17cm]{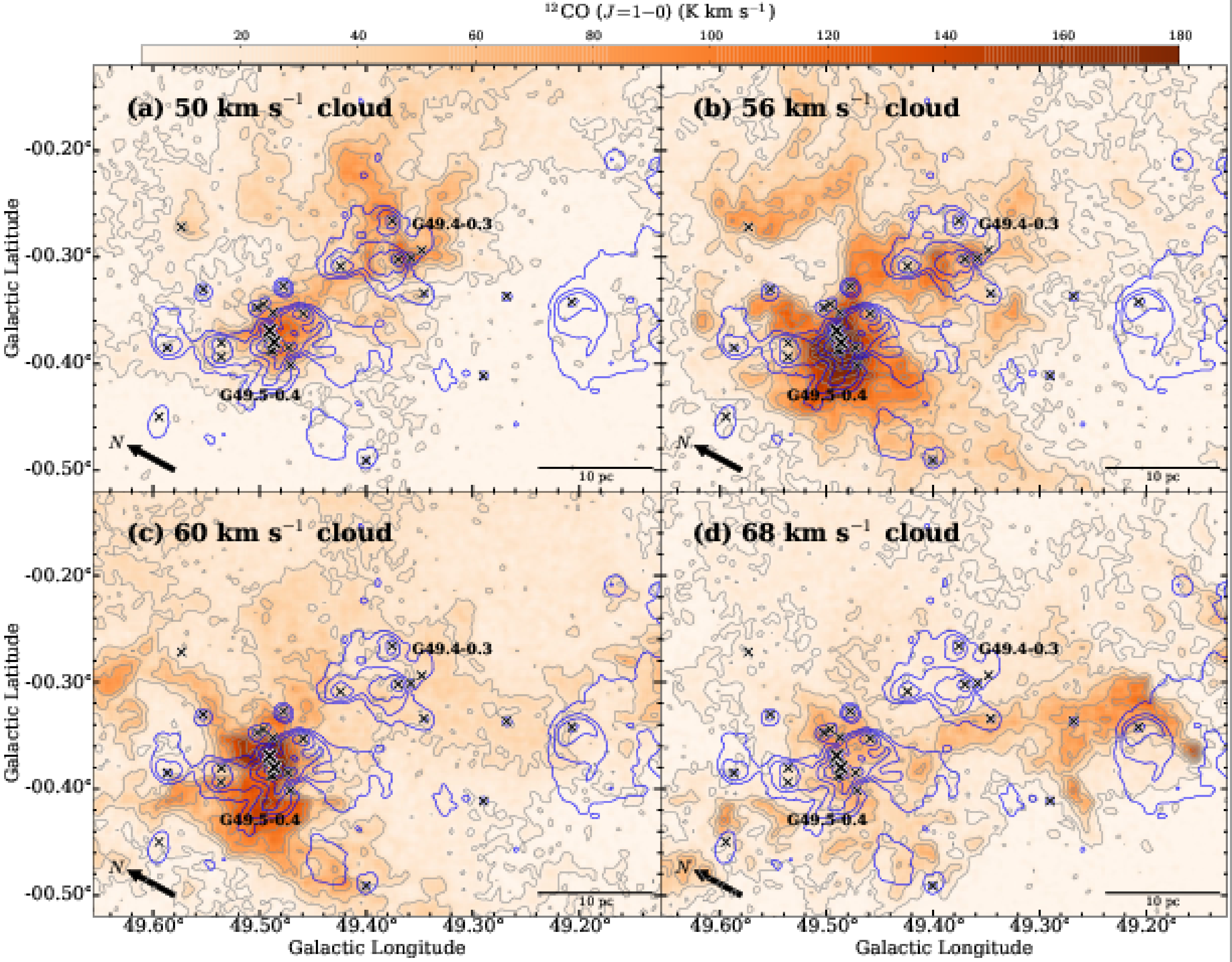}
  \end{center}
  \caption{The $^{12}$CO ($J$=1--0) integrated intensity distributions of the (a) 50, (b) 56, (c) 60, and (d) 68\,km\,s$^{-1}$ clouds, with the integration ranges of 46.9--52.1, 52.8--58.6, 59.3--64.5, and 65.1--71.0\,km\,s$^{-1}$, respectively. \textcolor{black}{The gray contours are plotted with $6\,\sigma$ (20\,K\,km\,s$^{-1}$) intervals starting from the $3\,\sigma$ (10\,K\,km\,s$^{-1}$) level. The blue contours show the THOR 21\,cm radio continuum emission combined with the VGPS data (\cite{2016A&A...595A..32B, 2006AJ....132.1158S}), and are plotted from 0.06 to 3.0\,Jy\,str$^{-1}$ with logarithm step. The angular resolution of the THOR data combined with the VGPS is 25$''$. } The crosses represent H{\sc ii} regions listed in \citet{1994ApJS...91..713M}.} \label{fig:four_integ_12}
\end{figure*}

\begin{figure*}[htbp]
  \begin{center}
         \includegraphics[width=17cm]{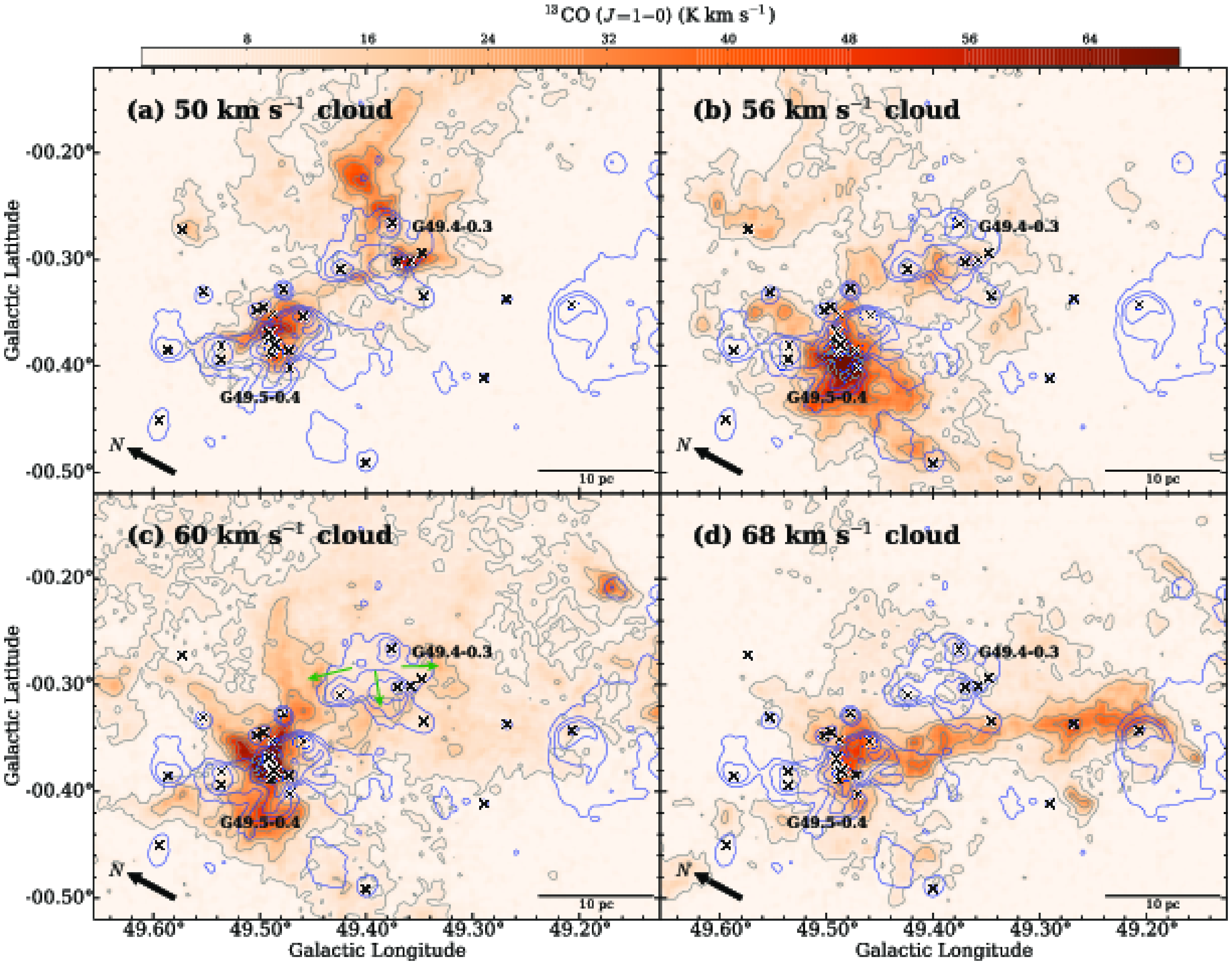}
  \end{center}
  \caption{Same as Figure\,\ref{fig:four_integ_12} but for the $^{13}$CO ($J$=1--0) emissions. \textcolor{black}{The gray contours are plotted with $6\,\sigma$ (10\,K\,km\,s$^{-1}$) intervals starting from the $3\,\sigma$ (5\,K\,km\,s$^{-1}$) level. }}\label{fig:four_integ_13}
\end{figure*}

\begin{figure*}[htbp]
  \begin{center}
         \includegraphics[width=17cm]{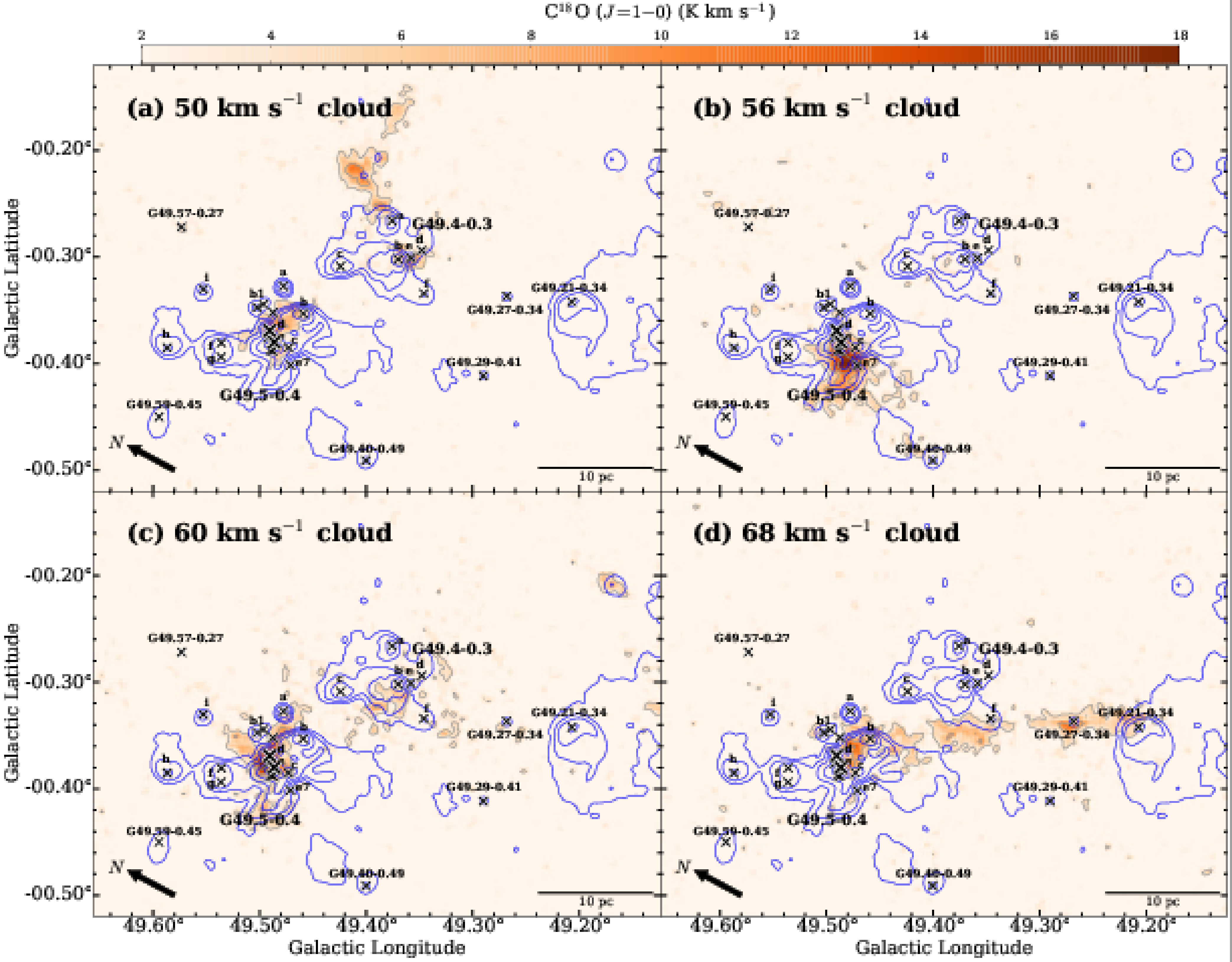}
  \end{center}
  \caption{Same as Figure\,\ref{fig:four_integ_12} but for the C$^{18}$O ($J$=1--0) emissions. \textcolor{black}{The gray contours are plotted with $3\,\sigma$ (4\,K\,km\,s$^{-1}$) intervals starting from the $3\,\sigma$ (4\,K\,km\,s$^{-1}$) level.}}\label{fig:four_integ_18}
\end{figure*}

Figures\,\ref{fig:four_integ_12}--\ref{fig:four_integ_18} show the $^{12}$CO, $^{13}$CO and C$^{18}$O ($J$=1--0) integrated intensity distributions of the four velocity clouds, respectively, overlaid with the contour map of the 21\,cm radio continuum data (\cite{2006AJ....132.1158S}).
We also present the velocity channel maps of the $^{12}$CO, $^{13}$CO, and C$^{18}$O emissions in Figure\,\ref{fig:chmap} in the appendix for supplements.
The $^{12}$CO emissions presented in Figure\,\ref{fig:four_integ_12} show extended gas distributions of the four clouds, while the $^{13}$CO emissions show that the clouds have networks of clumpy and filamentary structures.
C$^{18}$O probes only clumpy structures except for the 68\,km\,s$^{-1}$ cloud. 

The 50\,km\,s$^{-1}$ cloud shown in the panel (a) in Figures\,\ref{fig:four_integ_12}--\ref{fig:four_integ_18} has strong intensity peaks toward G49.5-0.4 and G49.4-0.3.
In the $^{13}$CO emissions these peaks are connected with each other with filamentary structures roughly elongated along the northeast-southwest at $b\sim-0\fdg38$--$-0\fdg32$ (Figure\,\ref{fig:four_integ_13}(a)), which are not apparent in the $^{12}$CO emissions (Figure\,\ref{fig:four_integ_12}(a)), whereas the C$^{18}$O emissions show fragmented distributions with sizes of $\sim$2--3\,pc toward these peaks (Figure\,\ref{fig:four_integ_18}(a)).
We derived the total molecular mass of the 50\,km\,s$^{-1}$ cloud as $\sim1.1\times10^5\ M_{\odot}$ using the $^{13}$CO ($J$=1--0) map in Figure\,\ref{fig:four_integ_13}(a) with an assumption of the local thermodynamic equilibrium (LTE) \textcolor{black}{(e.g., \cite{1998ApJS..117..387K})}.
We adopted an abundance ratio [$^{13}$CO]/[H$_2$] of $1.5 \times 10^{-6}$ (\cite{1978ApJS...37..407D}), \textcolor{black}{and we estimated excitation temperature $T_{\rm ex}$ for each pixel from the value of the peak brightness temperature of the optically thick $^{12}$CO ($J$=1--0) emissions in the 50\,km\,s$^{-1}$ cloud. }

The CO emissions in the 56\,km\,s$^{-1}$ cloud shown in the panel (b) in Figures\,\ref{fig:four_integ_12}--\ref{fig:four_integ_18} are enhanced at $(l,b)\sim(49\fdg48, -0\fdg40)$ in G49.5-0.4, whose CO intensities are strongest among the four velocity clouds in W51A.
The $^{13}$CO emissions show filamentary structures, and some of them are radially elongated from the CO peak at G49.5-0.4 (Figure\,\ref{fig:four_integ_13}(b)), while the C$^{18}$O emission is detected only toward the peak with a size of $\sim$3\,pc (Figure\,\ref{fig:four_integ_18}(b)).
The total molecular mass of the 56\,km\,s$^{-1}$ cloud measured using the $^{13}$CO map in Figure\,\ref{fig:four_integ_13}(b) is $\sim1.3\times10^5\ M_{\odot}$. 

The 60\,km\,s$^{-1}$ cloud in the panel (c) in Figures\,\ref{fig:four_integ_12}--\ref{fig:four_integ_18} shows similar gas distribution as the 56\,km\,s$^{-1}$ cloud, as in the $^{13}$CO emissions it consists of a strong CO peak at G49.5-0.4, attached with filamentary structures (Figure\,\ref{fig:four_integ_13}(c)). 
A difference between the 50 and 56\,km\,s$^{-1}$ clouds is the diffuse $^{13}$CO emissions extended above $b\sim-0\fdg3$ between $l$ of $49\fdg45$ and $49\fdg30$.
The total molecular mass of the 60\,km\,s$^{-1}$ cloud estimated with the $^{13}$CO map is as large as $\sim1.9\times10^5\ M_{\odot}$. 
The C$^{18}$O distribution in Figure\,\ref{fig:four_integ_18}(c) is highly fragmented in G49.5-0.4 and G49.4-0.3. 
Note that the C$^{18}$O fragments distributed to the east of G49.4-0.3 at $(l,b)\sim(49\fdg38, -0\fdg32)$ corresponds to a part of the $^{13}$CO filamentary structure which surrounds the 21\,cm contours of G49.4-0.3 (Figure\,\ref{fig:four_integ_13}(c)), suggesting possible interaction between the 60\,km\,s$^{-1}$ cloud and G49.4-0.3

In panel (d) in Figures\,\ref{fig:four_integ_12}--\ref{fig:four_integ_18}, the 68\,km\,s$^{-1}$ cloud has a filamentary structure elongated nearly parallel to the galactic plane between $l$ of $49\fdg50$ and $49\fdg20$. 
The width of the filament can be measured as $\sim$4--8\,pc in the $^{12}$CO and $^{13}$CO emissions, while it is thinner in the C$^{18}$O emissions, $\sim$3\,pc.
The northeastern tip of the filament is spatially coincident with G49.5-0.4, while the opposite end corresponds to the H{\sc II} region G49.21-0.34.
The total mass of the 68\,km\,s$^{-1}$ cloud is estimated as $\sim1.3\times10^5\,M_{\odot}$, which is consistent with the estimate in \citet{1998AJ....116.1856C}.

\subsubsection{CO ($J$=3--2/$J$=1--0) intensity ratios}

\begin{figure*}[htbp]
  \begin{center}
         \includegraphics[width=12cm]{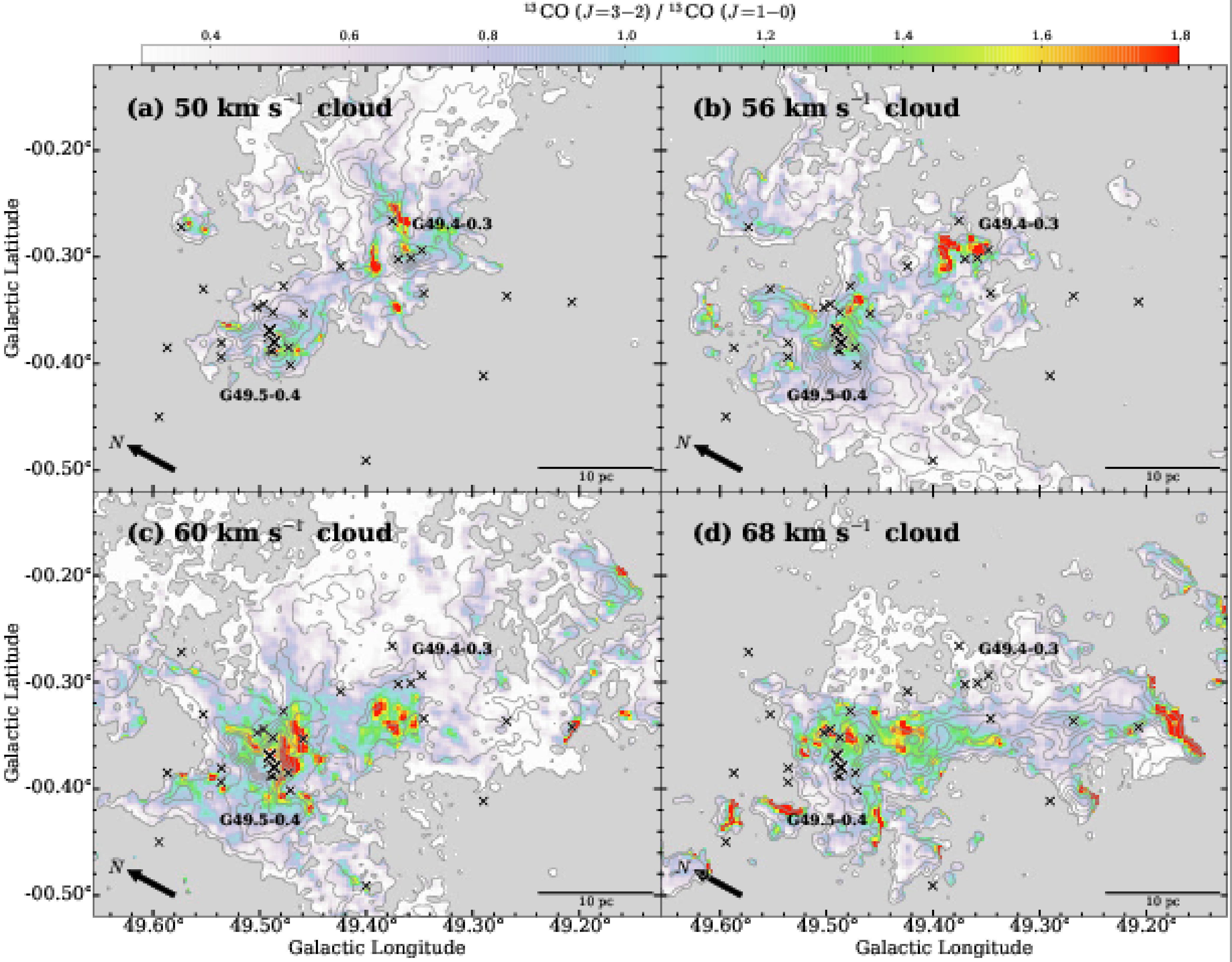}
         \includegraphics[width=12cm]{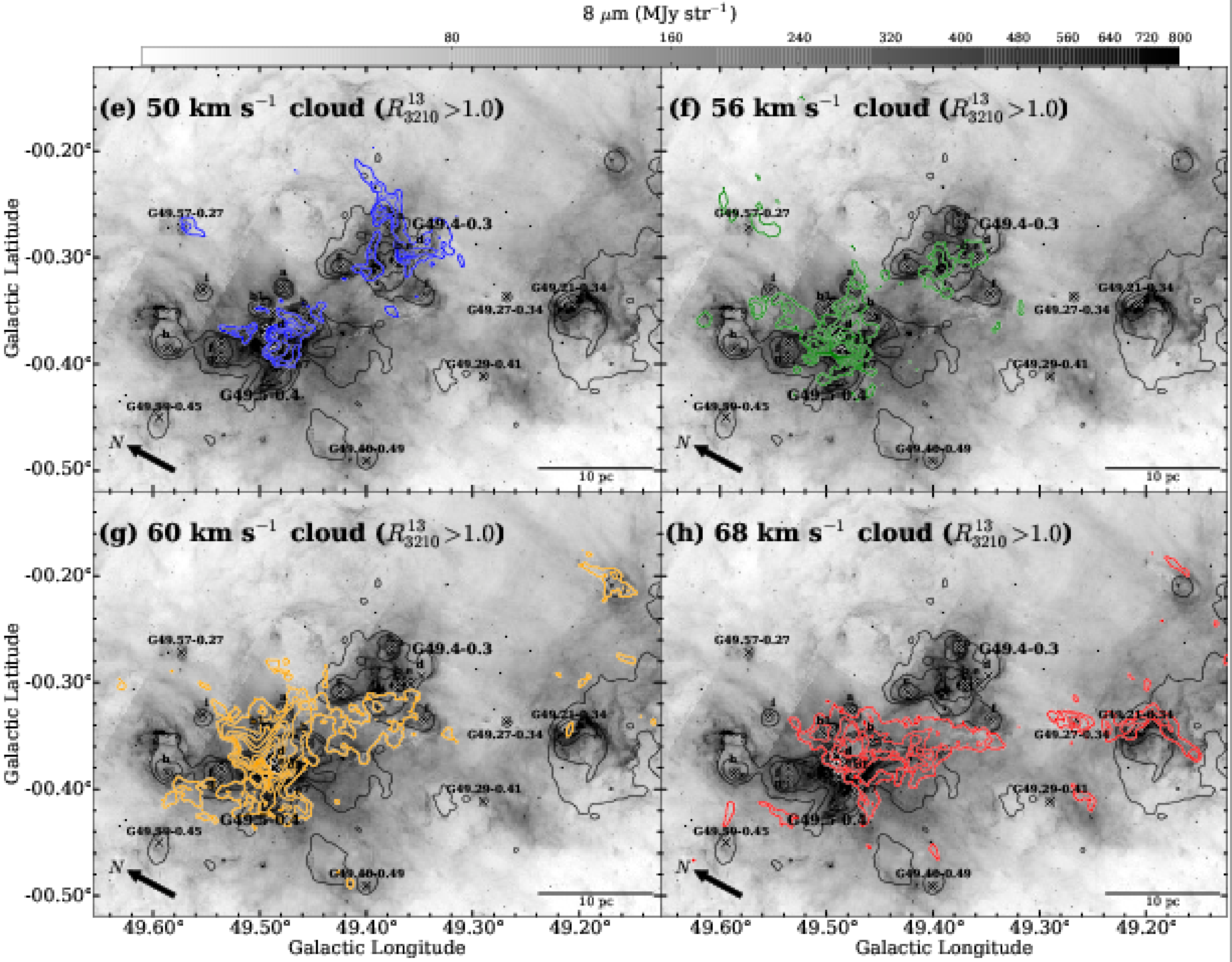}
  \end{center}
  \caption{(a--d) The $R_{3210}^{13}$ distributions of the four clouds. The contours indicate $^{13}$CO ($J$=1--0), and are plotted with $6\,\sigma$ (10\,K\,km\,s$^{-1}$) intervals starting from the $6\,\sigma$ (10\,K\,km\,s$^{-1}$) level for (a)--(c), and $6\,\sigma$ (5\,K\,km\,s$^{-1}$) intervals starting from the $5\,\sigma$ (8\,K\,km\,s$^{-1}$) level for (d), respectively. \textcolor{black}{The associated errors of $R_{3210}^{13}$ for each pixels are presented in Figure\,\ref{fig:ratio_error} in Appendix\,\ref{app:ratio_error}.} (e--h) Spatial distributions of the high-$R_{3210}^{13}$ gas are shown in the colored contour maps, where the high-$R_{3210}^{13}$ data was made by integrating only the voxels having $R_{3210}^{13} > 1.0$. The colored contours are plotted at the same levels as in (a)--(d). The background image indicates the {\it Spitzer} 8\,$\mu$m image, while the black contours represent the 21\,cm continuum emissions plotted at the same levels as those in Figure\,\ref{fig:four_integ_12}. Crosses are the same as those in Figure\,\ref{fig:four_integ_12}. }\label{fig:four_R3210}
\end{figure*}

Figures\,\ref{fig:four_R3210}(a)--(d) show large-scale distributions of the $^{13}$CO ($J$=3--2)/$^{13}$CO ($J$=1--0) \textcolor{black}{integrated} intensity ratios (hereafter $R_{3210}^{13}$) of the four clouds in W51A \textcolor{black}{(The associated errors of $R_{3210}^{13}$ is presented in Appendix\,\ref{app:ratio_error})}. 
The four clouds typically have $R_{3210}^{13}$ of higher than 0.6, up to over 2.0, while low $R_{3210}^{13}$ less than 0.2 is seen in the diffuse gas widely distributed at $b > -0\fdg30$--$-0\fdg20$ and $b < -0\fdg45$.
Moreover, to extract the molecular gas heated up by the massive stars in W51A, in Figures\,\ref{fig:four_R3210}(e)--(h) we plot the $^{13}$CO ($J$=1--0) contour maps of the four clouds using only the voxels having $R_{3210}^{13}$ of higher than 1.0.
In Appendix\,\ref{app:LVG}, we performed the large velocity gradient (LVG) analysis, indicating that $R_{3210}^{13}$ of higher than 1.0 can probe the high-temperature gas having $>20$\,K.

In G49.5-0.4, the high-$R_{3210}^{13}$ gas is seen in all four clouds, and physical associations between these clouds and the H{\sc ii} regions in G49.5-0.4 are thus suggested.
In G49.4-0.3, on the other hand, the filamentary structures seen in the 8\,$\mu$m image are well traced by the high-$R_{3210}^{13}$ gas in the 50\,km\,s$^{-1}$ cloud, forming an arch-like gas distribution (Figures\,\ref{fig:four_R3210}(a) and (e)). 
The 56\,km\,s$^{-1}$ cloud is also high in $R_{3210}^{13}$ at the footpoints of the arch-like structure (Figures\,\ref{fig:four_R3210}(b) and (f)).
In the 60\,km\,s$^{-1}$ cloud, high-$R_{3210}^{13}$ gas look surrounding the eastern and southeastern rim of the 21\,cm emissions of  G49.4-0.3 (Figures\,\ref{fig:four_R3210}(c) and (g)). 
The 68\,km\,s$^{-1}$ cloud show continuous distribution of high-$R_{3210}^{13}$ gas between the east and the south of G49.4-0.3, although its association with G49.4-0.3 is not clear (Figures\,\ref{fig:four_R3210}(d) and (h)).
These results indicate physical associations of multiple velocity components of gas with G49.5-0.4 and G49.4-0.3.

Associations of the other H{\sc ii} regions with the multiple velocity components can also be investigated in Figures\,\ref{fig:four_R3210}(e)--(h).
G49.57-0.27 is an isolated H{\sc ii} region situated at the north of G49.5-0.4, which shows high $R_{3210}^{13}$ in the 50 and 56\,km\,s$^{-1}$ clouds (Figures\,\ref{fig:four_R3210}(e) and (f)).
There are several relatively expanded H{\sc ii} regions in G49.5-0.4, i.e., G49.5-0.4f, g, h, and i.
Although we found no high-$R_{3210}^{13}$ gas spatially overlapping these H{\sc ii} regions, there are several high-$R_{3210}^{13}$ components in the 50, 56, and 60,\,km\,s$^{-1}$ clouds which are distributed at the rims of the 21\,cm continuum emissions of these H{\sc ii} regions (Figures\,\ref{fig:four_R3210}(e)--(g)).
As these four H{\sc ii} regions are relatively evolved with ages of an order of 1\,Myr as summarized in Table\,\ref{tab:sources}, these high-$R_{3210}^{13}$ gas can be interpreted as the remnants of the natal molecular gas of the massive stars in these H{\sc ii} regions.
In the other H{\sc ii} regions, we found no plausible signatures of physical association of multiple velocity components. These show either no high-$R_{3210}^{13}$ gas in the four clouds (e.g., G49.29-0.41) or high-$R_{3210}^{13}$ gas only in one cloud (e.g., G49.21-0.34, G49.27-0.34, and G49.59-0.45).
In the next subsection, we present the detailed gas distribution toward the individual H{\sc ii} regions which are likely associated with multiple velocity clouds.

\subsection{Detailed gas distributions toward individual H{\sc ii} regions}
\subsubsection{G49.5-0.4}

\begin{figure}[h]
 \begin{center}
  \includegraphics[width=15cm]{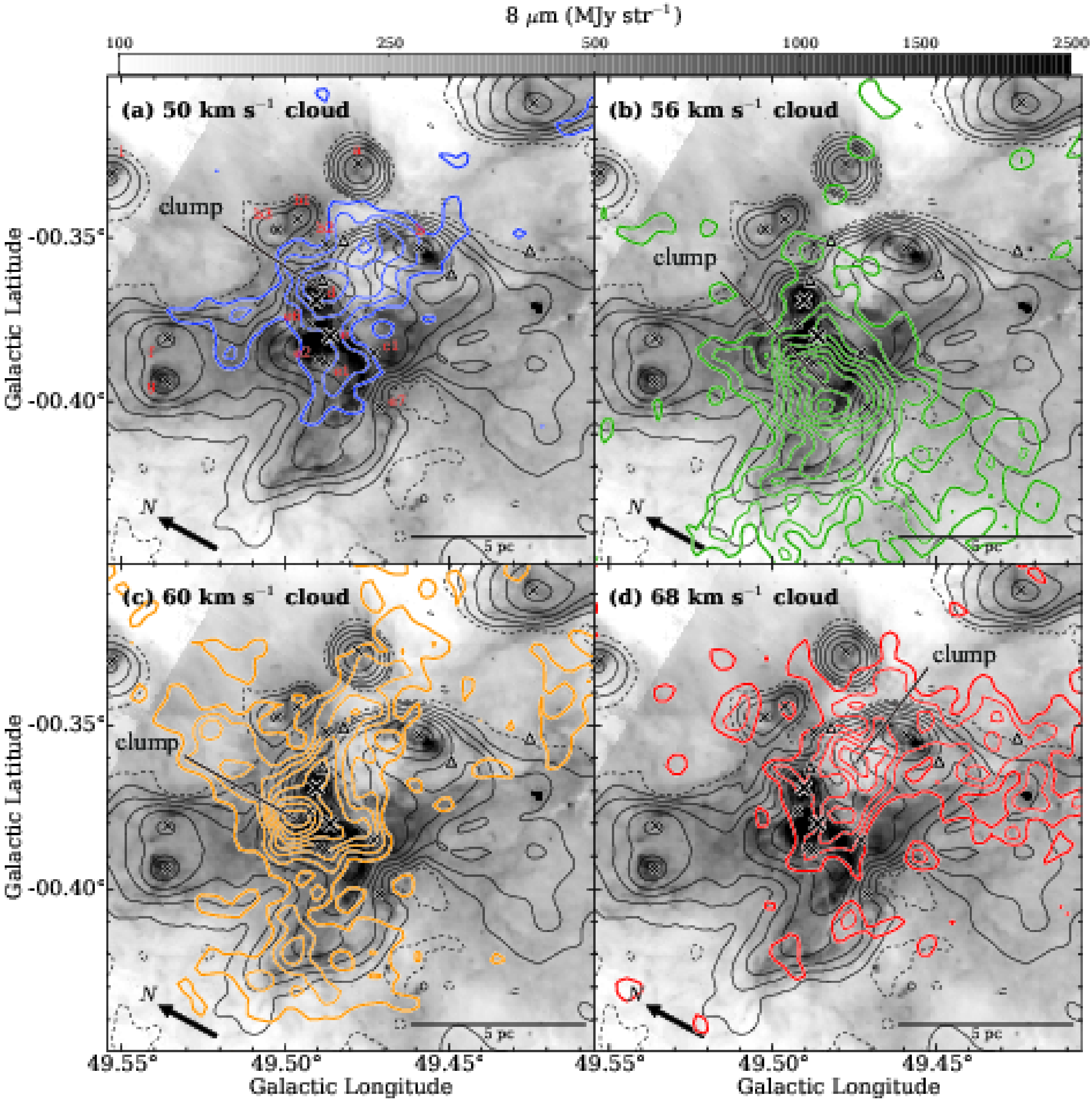}
 \end{center}
 \caption{C$^{18}$O ($J$=1--0) integrated intensity distributions of the four velocity clouds in G49.5-0.4 are presented as colored contour maps superimposed on the {\it Spitzer} 8\,$\mu$m image (\cite{2009PASP..121...76C}). The colored contours are plotted with 2\,K\,km\,s$^{-1}$ intervals starting from the 3\,K\,km\,s$^{-1}$ level where the RMS noise level of the image is $\sim$ 1.5\,K\,km\,s$^{-1}$. The H{\sc ii} regions listed in Table\,\ref{tab:sources} are depicted by crosses, and the massive young stars identified by \citet{2017ApJ...839..108S} are plotted with triangles. The black contours show the 21\,cm continuum emissions plotted from 0.03 (dashed lines) to 3.0\,Jy\,str$^{-1}$ with logarithm step. }\label{fig:comp8_c18o}
\end{figure}

Figure\,\ref{fig:comp8_c18o} shows a close-up view of G49.5-0.4 with comparisons between the {\it Spitzer} 8\,$\mu$m image \citep{2009PASP..121...76C} and the C$^{18}$O contour maps of the four clouds, where the H{\sc ii} regions listed in Table\,\ref{tab:sources} are depicted by crosses, and the MYSOs identified by \citet{2017ApJ...839..108S} are plotted with triangles.
The 8\,$\mu$m emission is bright around $(l,b)\sim(49\fdg45$--$49\fdg49$, $-0\fdg40$--$-0\fdg33)$, at which many compact H{\sc ii} regions including IRS\,1 (G49.5-0.4e) and IRS\,2 (G49.5-0.4d) are concentrated.
Figure\,\ref{fig:comp8_c18o} shows that the four clouds have compact and bright C$^{18}$O emissions within a few pc of the central 8\,$\mu$m structure.

\begin{figure}[htbp]
  \begin{center}
          \includegraphics[width=8cm]{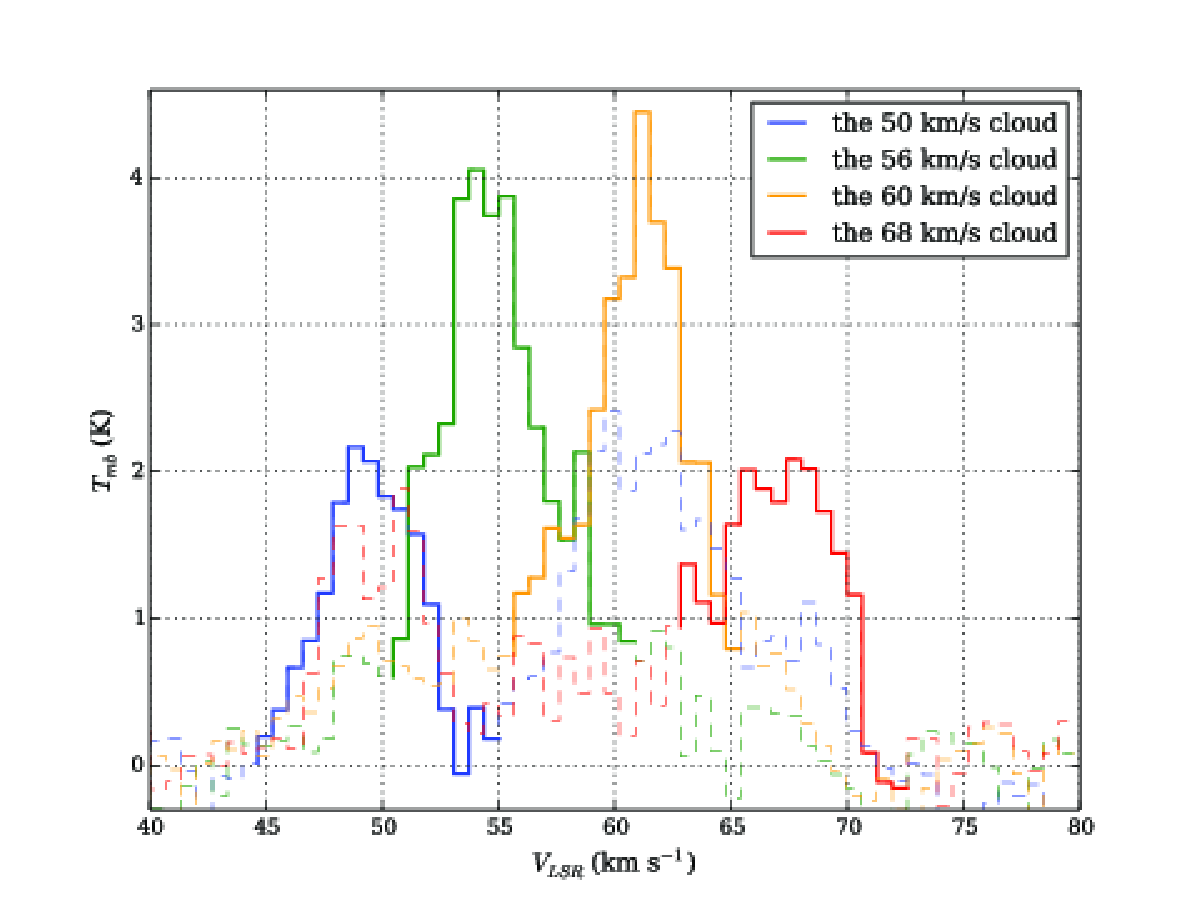}
  \end{center}
  \caption{The C$^{18}$O ($J$=1--0) spectra at the peak position (see Table\,\ref{tab:C18O}) of each the clump embedded in the 50\,km\,s$^{-1}$ (blue), the 56\,km\,s$^{-1}$ (green), the 60\,km\,s$^{-1}$ (orange), and the 68\,km\,s$^{-1}$ (red) cloud. \textcolor{black}{Velocities outside $\pm \,5$\,km\,s$^{-1}$ of each peak of the clump are plotted with dashed lines.}}\label{fig:spe_C18O_G49.5-0.4}
\end{figure}

We identified C$^{18}$O clumps in this region by drawing a contour at the 70\% level of the maximum integrated intensity in the four clouds, resulted in discoveries of the four C$^{18}$O clumps which are each embedded within the four clouds. 
\textcolor{black}{
Figure\,\ref{fig:spe_C18O_G49.5-0.4} shows the C$^{18}$O ($J$=1--0) spectra at the peak position of each clump. 
\textcolor{black}{Velocities outside $\pm \,5$\,km\,s$^{-1}$ of each peak of the clump are plotted with dashed lines.}
The physical parameters are summarized in Table\,\ref{tab:C18O}.
The FWHM velocity widths of each spectra are approximately 5--6\,km\,s$^{-1}$. 
}
The peak column densities of the four clumps in the 50, 56, 60, and 68\,km\,s$^{-1}$ clouds are measured as 2.3, 4.4, 3.9, and 2.7\,$(\times10^{23})$\,cm$^{-2}$, respectively, from the C$^{18}$O ($J$=1--0) data by assuming LTE. 
We here adopted an abundance ratio of [C$^{18}$O]/[H$_2$]\,=\,$1.7\times10^{-7}$ \citep{1982ApJ...262..590F}.
\textcolor{black}{
If we tentatively determine the radii of the clumps as 70\% of the peak, virial masses are estimated to be Table\,\ref{tab:C18O}(8). 
The LTE masses (Table\,\ref{tab:C18O}(5)) are almost greater than the virial masses. 
}

\begin{figure}[htbp]
  \begin{center}
          \includegraphics[width=17cm]{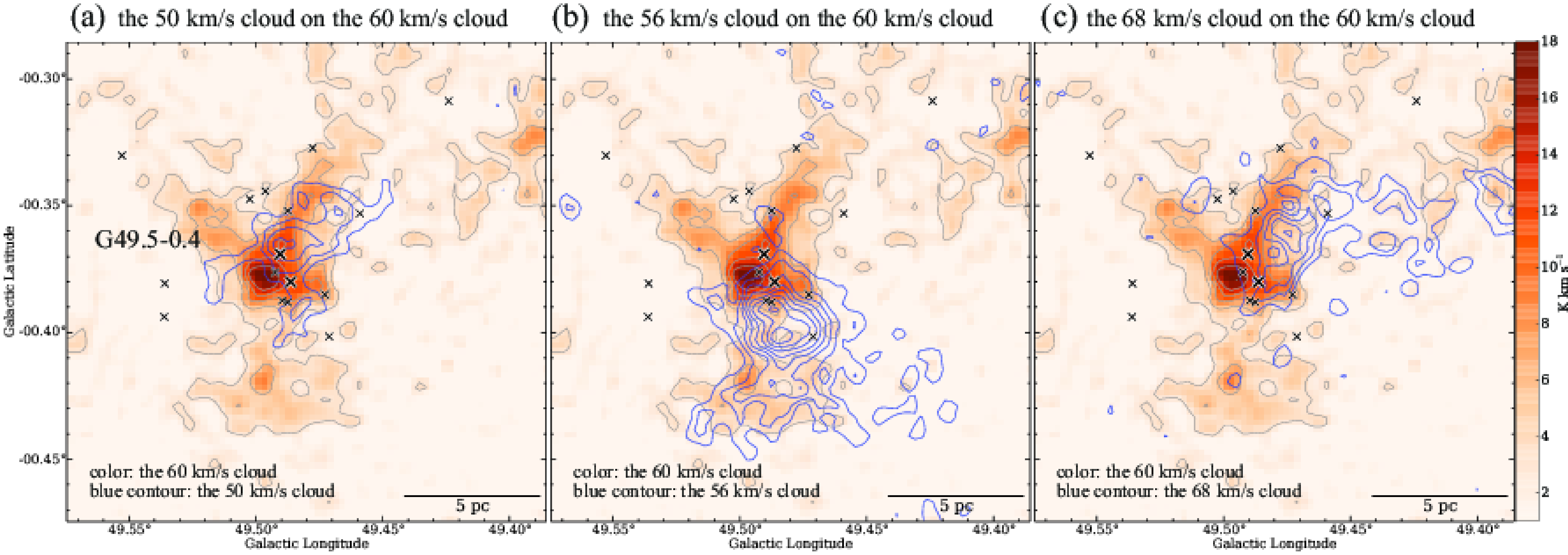}
  \end{center}
  \caption{\textcolor{black}{(a) The C$^{18}$O ($J$=1--0) integrated intensity distributions of the 60\,km\,s$^{-1}$ cloud. Overlaid blue contours show the C$^{18}$O ($J$=1--0) integrated intensity distributions of the 50\,km\,s$^{-1}$ cloud, and are plotted with $1.5\,\sigma$ (2\,K\,km\,s$^{-1}$) intervals starting from the $3\,\sigma$ (4\,K\,km\,s$^{-1}$) level. The H{\sc ii} regions listed in Table\,\ref{tab:sources} are denoted by crosses. (b) The same as (a), but the blue contours show the C$^{18}$O ($J$=1--0) integrated intensity distributions of the 56\,km\,s$^{-1}$ cloud.  (c) The same as (a), but the blue contours show the C$^{18}$O ($J$=1--0) integrated intensity distributions of the 68\,km\,s$^{-1}$ cloud. }}\label{fig:C18O_comp}
\end{figure}

\textcolor{black}{
Color scale in Figures\,\ref{fig:C18O_comp}(a), \ref{fig:C18O_comp}(b), and \ref{fig:C18O_comp}(c) show the C$^{18}$O ($J$=1--0) integrated intensity distributions of the 60\,km\,s$^{-1}$ cloud, and the blue contours show the C$^{18}$O ($J$=1--0) integrated intensity distributions of (a) the 50\,km\,s$^{-1}$ cloud, (b) the 56\,km\,s$^{-1}$ cloud, and (c) the 68\,km\,s$^{-1}$ cloud, respectively. 
In Figure\,\ref{fig:C18O_comp}(a), the 60\,km\,s$^{-1}$ cloud surrounds the peak of the 50\,km\,s$^{-1}$ cloud. 
In Figure\,\ref{fig:C18O_comp}(b), the 56\,km\,s$^{-1}$ cloud and the 60\,km\,s$^{-1}$ cloud show a complementary distribution. 
Futhermore, in Figure\,\ref{fig:C18O_comp}(c), the 60\,km\,s$^{-1}$ cloud and the 68\,km\,s$^{-1}$ cloud also show a complementary distribution. 
The complementary distribution between the 60 and 68\,km\,s$^{-1}$ clouds was discussed by \citet{1998AJ....116.1856C} based on their CO observations, and a CCC scenario between these two clouds were suggested by the authors.
}

Figure\,\ref{fig:G49.5-0.4_vb}(a) shows comparisons of the identified four C$^{18}$O clumps superimposed on the {\it Spitzer} 8\,$\mu$m image, where the C$^{18}$O ($J$=1--0) contours are plotted at 60, 70, 80, and 90\,\% of the peak intensities of the clumps.
Although these four clumps are concentrated within a small area of less than 5\,pc, these are not spatially coincident along the line of sight, showing a complementary distribution.
Our CO data newly revealed that the complementary distribution are seen not only for the 60 and 68\,km\,s$^{-1}$ clouds, but also for all four clouds. 

The compact H{\sc ii} regions depicted by crosses are distributed around the rims of these four clumps.
IRS\,2 (G49.5-0.4d) is distributed at the interface of the clumps between the 50 and 60\,km\,s$^{-1}$ clouds, while IRS\,1 (G49.5-0.4e) is seen at the boundaries of the 56, 60, and 68\,km\,s$^{-1}$ clouds.
The other H{\sc ii} regions G49.5-0.4e1, e2, and e6 also are distributed at the interface of the four clouds, where the 8\,$\mu$m emission is also enhanced (Figure\,\ref{fig:G49.5-0.4_vb}(a)).
The $v$-$b$ diagram of the $^{13}$CO and C$^{18}$O ($J$=1--0) emissions shown in Figures\,\ref{fig:G49.5-0.4_vb}(b) indicates that the four clumps are connected with each other by the CO emissions with intermediate intensities.
These intermediate velocity features are possibly interpreted as the broad bridge features created in the CCC process as discussed in Subsection\,1.2.
\textcolor{black}{
\textcolor{black}{Figure\,}\ref{fig:G49.5-0.4_vb}(c) shows the $v$-$b$ diagram of the $R_{3210}^{13}$. 
The high $R_{3210}^{13}$ \textcolor{black}{($>\,1.5$)} are seen at the velocity edge and the intermediate velocity of the clouds.
\textcolor{black}{In the central molecular zone of the Galaxy, where is a spectacular star-forming environment, \citet{2007PASJ...59...15O} also reported a high CO $J$=3--2/$J$=1--0 ratio of $>\,1.5$.}
The high $R_{3210}^{13}$ in G49.5-0.4 are probably caused by the feedback from the massive stars and/or collisional heating between the clouds. 
}

\begin{figure}[htbp]
  \begin{center}
          \includegraphics[width=17cm]{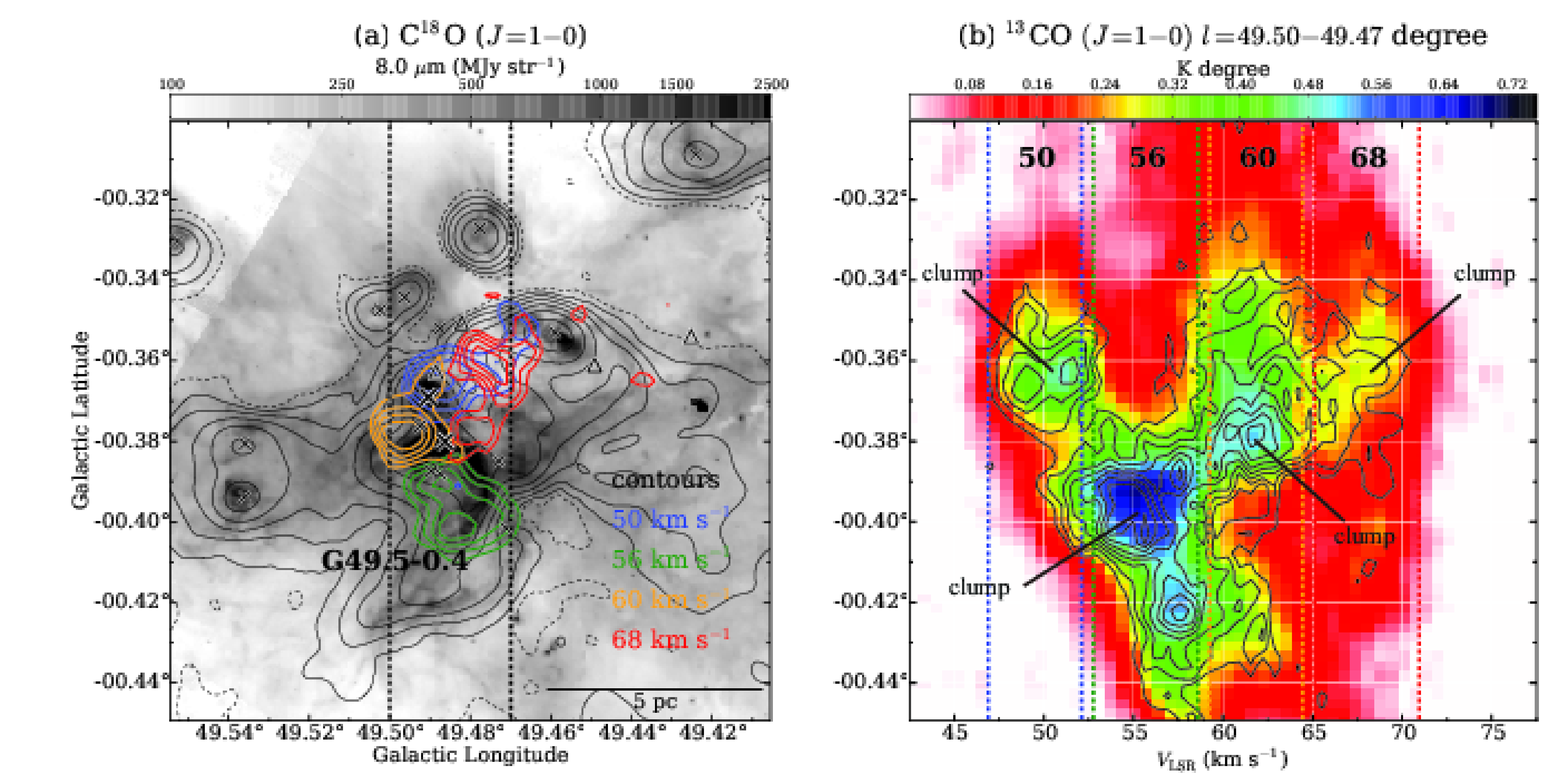}
  \end{center}
  \caption{(a) The contour maps of the C$^{18}$O ($J$=1--0) emissions of the four clouds around G49.5-0.4 are plotted superimposed on the {\it Spitzer} 8$\mu$m image and 21\,cm contour map. The C$^{18}$O ($J$=1--0) contours are plotted at 60, 70, 80, and 90 \% of the peak intensities of the four clumps. The black contours show the 21\,cm continuum emissions plotted from 0.03 (dashed lines) to 3.0\,Jy\,str$^{-1}$ with logarithm step. (b) Velocity--Galactic Latitude ($v$--$b$) diagram of the $^{13}$CO ($J$=1--0) (color) and C$^{18}$O ($J$=1--0) emissions (contours) towards G49.5-0.4 integrated over $l=49\fdg 50$--$49\fdg 47$, where the contours are plotted with $5\,\sigma$ (0.03\,K\,degrees) intervals starting from the $5\,\sigma$ (0.03\,K\,degrees) level. The vertical dashed lines show the integration ranges of the four clouds presented in Figures\,\ref{fig:comp8_c18o} and \ref{fig:G49.5-0.4_vb}(a).}\label{fig:G49.5-0.4_vb}
\end{figure}

\begin{figure}[htbp]
  \begin{center}
          \includegraphics[width=8cm]{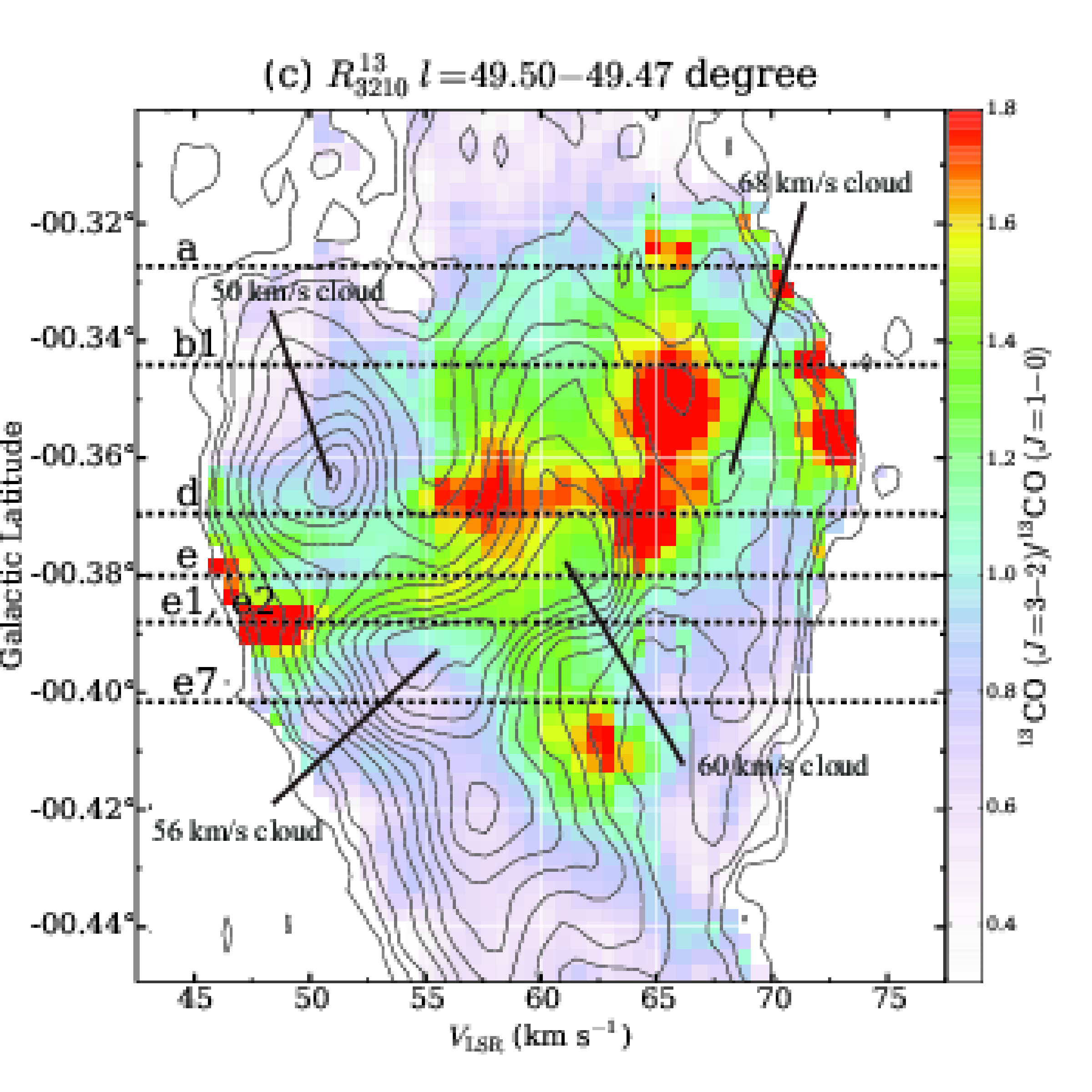}
  \end{center}
  \contcaption{\textcolor{black}{Continued. (c) The $v$--$b$ diagram of the $R_{3210}^{13}$ distributions toward G49.5-0.4 integrated over $l=49\fdg 50$--$49\fdg 47$. The contours show the intensity of the $^{13}$CO ($J$=1--0). The horizontal dashed lines indicate the position of the compact H{\sc ii} regions G49.5-0.4a, b, d, e, e1, e2, and e7. }}
\end{figure}

 \begin{table}[h]
   \tbl{Parameters of the C$^{18}$O ($J$=1--0) clumps in G49.5-0.4 \textcolor{black}{and G49.4-0.3} }
   {%
    \begin{tabular}{cccccccc}
    \hline \hline
    cloud name & peak position & \textcolor{black}{C$^{18}$O (1--0) $W_{\rm max}$} & \textcolor{black}{$R\, {\rm (70\%)}$} & \textcolor{black}{$M_{\rm LTE}\, {\rm (70\%)}$} &$N_{\rm max}({\rm H_2})$  &  \textcolor{black}{FWHM} &  \textcolor{black}{$M_{\rm vir}$} \\
       & $(l,b)$ & (K\,km\,s$^{-1}$) & (pc) & ($10^4\ M_{\odot}$) & ($10^{23}\ {\rm cm^{-2}}$) & (km\,s$^{-1}$) & ($10^4\ M_{\odot}$)\\
    (1) & (2) & (3) & (4) & (5) & (6) & (7) & (8) \\   
       \hline \hline
       G49.5-0.4 & & & & & & \\
       50\,km\,s$^{-1}$ cloud & $49\fdg492, -0\fdg365$ & 11.0 & $\sim$\,1.1 & 0.9$\pm$0.1 & 2.3$\pm$0.2 & 4.7$\pm$0.2 & 0.6$\pm$0.1 \\
       56\,km\,s$^{-1}$ cloud & $49\fdg485, -0\fdg400$ & 19.7 & $\sim$\,1.1 & 2.3$\pm$0.1 & 4.4$\pm$0.4 & 6.3$\pm$0.5 & 1.0$\pm$0.2  \\
       60\,km\,s$^{-1}$ cloud & $49\fdg495, -0\fdg379$ & 18.0 & $\sim$\,0.8 & 1.0$\pm$0.1 & 3.9$\pm$0.4 & 6.0$\pm$0.5 & 0.7$\pm$0.1  \\
       68\,km\,s$^{-1}$ cloud & $49\fdg473, -0\fdg360$ & 12.2 & $\sim$\,1.0 & 1.4$\pm$0.1 & 2.7$\pm$0.3 & 6.4$\pm$0.7 & 1.0$\pm$0.2  \\ \hline
       \textcolor{black}{G49.4-0.3} & & & & & & &  \\
       50\,km\,s$^{-1}$ cloud & $49\fdg360, -0\fdg303$ & 13.3 & $\sim$\,0.8 & 0.3$\pm$0.1 & 2.1$\pm$0.2 & 3.2$\pm$0.1 & 0.2$\pm$0.1  \\
       56\,km\,s$^{-1}$ cloud & -- & -- & -- & -- & -- & -- & -- \\
       60\,km\,s$^{-1}$ cloud & $49\fdg391, -0\fdg322$ & 8.5 & $\sim$\,0.9 & 0.4$\pm$0.1 & 1.8$\pm$0.2 & 4.2$\pm$0.5 & 0.4$\pm$0.1  \\
       68\,km\,s$^{-1}$ cloud & $49\fdg355, -0\fdg355$ & 10.2 & $\sim$\,2.5 $\times$1.0 & 1.6$\pm$0.2 & 2.1$\pm$0.2 & 4.2$\pm$0.3 & 0.7$\pm$0.1  \\
       \hline
    \end{tabular}} \label{tab:C18O}
\begin{tabnote}
(1) Name of the cloud in which the clump is embedded. 
(2) Peak position of the C$^{18}$O ($J$=1--0) integrated intensity. 
(3) Peak integrated intensity the C$^{18}$O ($J$=1--0). 
(4) \textcolor{black}{Effective} radius of the clump, which was measured at $70\%$ of the peak intensity of the clump. 
(5) Molecular mass derived from C$^{18}$O ($J$=1--0) intensity maps assuming LTE \textcolor{black}{within the 70\% radius}. 
(6) Maximum H$_2$ column density of the clump. 
\textcolor{black}{(7) FWHM of the C$^{18}$O ($J$=1--0) spectrum at the peak position. These are derived by Gaussian fitting. }
\textcolor{black}{(8) Virial mass derived from column (4) and (7); $R\Delta v^2/G$ \textcolor{black}{, where $\Delta v$ and $G$ are the FWHM and the gravitational constant, respectively}}.
\end{tabnote}
\end{table}

\subsubsection{G49.4-0.3}

\begin{figure}[h]
 \begin{center}
  \includegraphics[width=15cm]{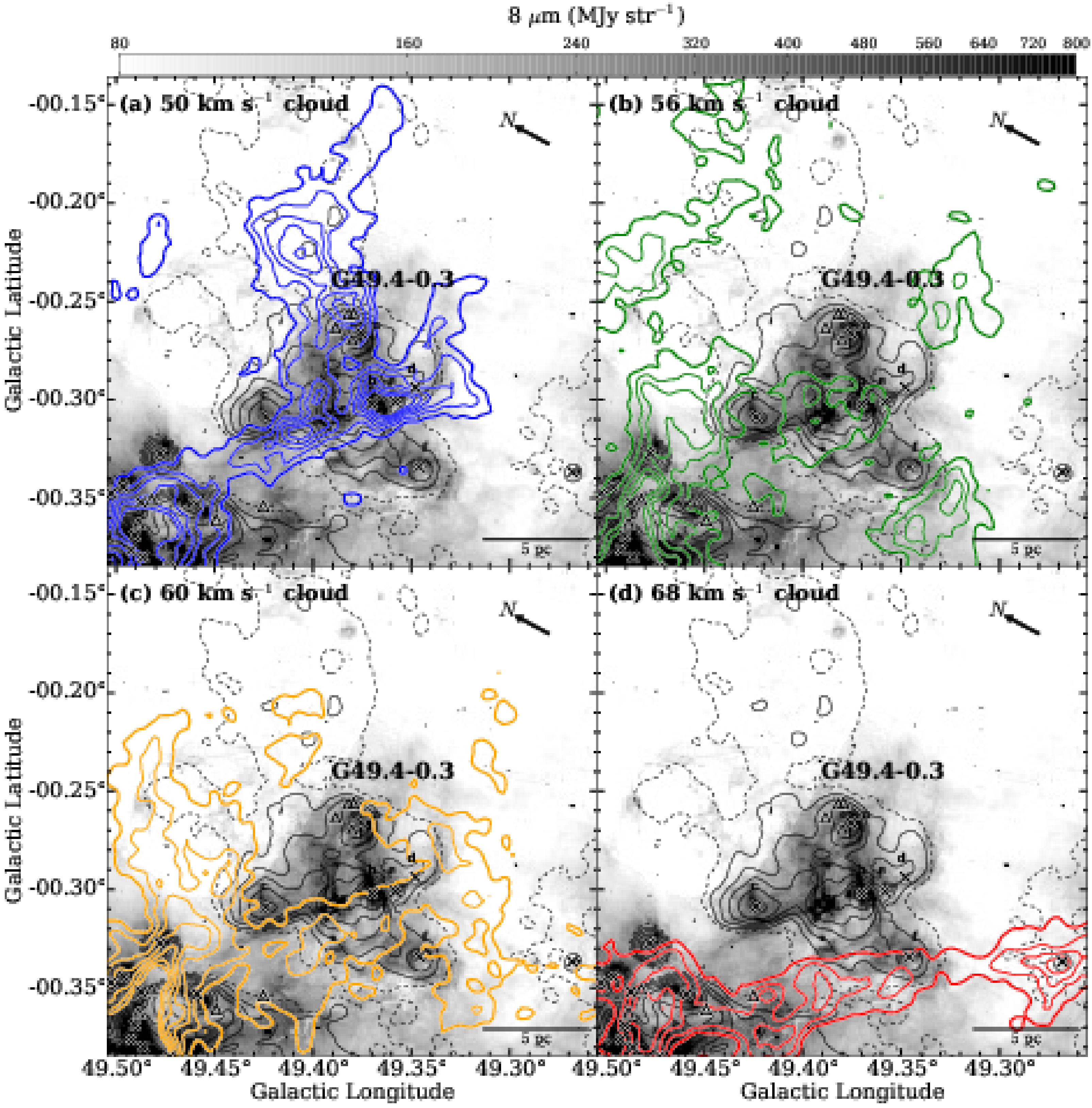}
 \end{center}
 \caption{$^{13}$CO ($J$=1--0) integrated intensity distributions of the four velocity clouds in G49.4-0.3 are presented as colored contour maps superimposed on the {\it Spitzer} 8\,$\mu$m image (\cite{2009PASP..121...76C}). Velocity ranges of the $^{13}$CO emissions in the panels (a)--(d) are 46.9--55.4, 55.4--59.3, 59.3--64.5, and 64.5--71.0\,km\,s$^{-1}$, and contours are plotted with 8\,K\,km\,s$^{-1}$ intervals starting from the 18\,K\,km\,s$^{-1}$ level for (a), 4\,K\,km\,s$^{-1}$ intervals starting from the 4\,K\,km\,s$^{-1}$ level for (b), 6\,K\,km\,s$^{-1}$ intervals starting from the 12\,K\,km\,s$^{-1}$ level for (c), and 8\,K\,km\,s$^{-1}$ intervals starting from the 12\,K\,km\,s$^{-1}$ level for (d), respectively, where the RMS noise level of the images are $\sim$1.5\,K\,km\,s$^{-1}$. Black contours show the 21\,cm emissions plotted at the same levels as those in Figure\,\ref{fig:four_integ_12}. The H{\sc ii} regions listed in Table\,\ref{tab:sources} are depicted by crosses, and the MYSOs identified by \citet{2017ApJ...839..108S} are plotted with triangles.}\label{fig:comp8_13co_494-03}
\end{figure}

\begin{figure*}[htbp]
  \begin{center}
         \includegraphics[width=17cm]{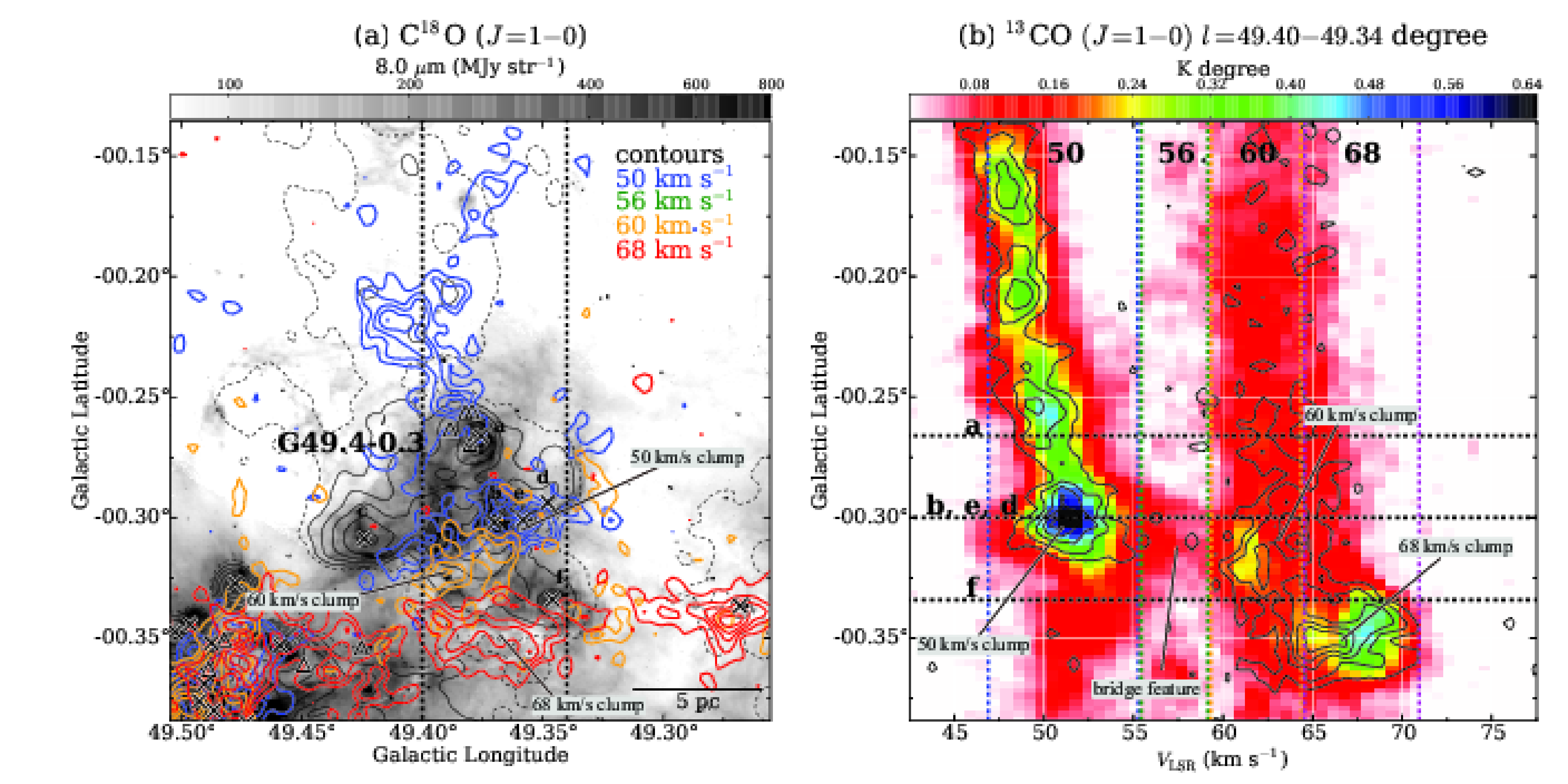}
  \end{center}
  \caption{(a) The colored contour maps (blue, green, orange, and red) of the C$^{18}$O emissions of the four clouds around G49.4-0.3 are plotted superimposed on the {\it Spitzer} 8$\mu$m image and 21\,cm contour map (black). C$^{18}$O contours are plotted with 2\,K\,km\,s$^{-1}$ intervals starting from the $3\,\sigma$ (4\,K\,km\,s$^{-1}$) level. Crosses represent compact H{\scriptsize II} regions listed in \citet{1994ApJS...91..713M} (Table\,\ref{tab:sources}), while the MYSO identified by \citet{2017ApJ...839..108S} is plotted with triangle. (b) Velocity--Galactic Latitude ($v$--$b$) diagram of $^{13}$CO ($J$=1--0) integrated over $l=49\fdg 40$--$49\fdg 34$. Contours indicate the C$^{18}$O ($J$=1--0) emissions and are plotted with $3\,\sigma$ (0.02\,K\,degrees) intervals starting from the $3\,\sigma$ (0.02\,K\,degrees) level. Horizontal dashed lines indicate the positions of compact H{\sc ii} regions, while the vertical dashed lines show the integration ranges of the four clouds presented in Figures\,\ref{fig:comp8_13co_494-03} and \ref{fig:G49.4-0.3_pv}(a). }\label{fig:G49.4-0.3_pv}
\end{figure*}

Figure\,\ref{fig:comp8_13co_494-03} shows the $^{13}$CO ($J$=1--0) contour maps of the four clouds toward G49.4-0.3.
The velocity ranges of the four clouds in each panel are determined from the $v$-$b$ diagram plotted in Figure\,\ref{fig:G49.4-0.3_pv}(b).
As already presented in Figure\,\ref{fig:four_R3210}, the arch-like CO structure in the 50\,km\,s$^{-1}$ cloud in Figure\,\ref{fig:comp8_13co_494-03}(a) is spatially correlated with the bright 8\,$\mu$m emissions, which consist of a network of the filamentary structures elongated nearly parallel or perpendicular to the galactic plane.
The 8\,$\mu$m filaments include the H{\sc ii} regions G49.4-0.3a, b, c, d, and e (see Table\,\ref{tab:sources}) and MYSOs. 
G49.4-0.3a, b and e out of them are spatially coincident with the 50\,km\,s$^{-1}$ cloud, as discussed by \citet{2010ApJS..190...58K}.

The 56\,km\,s$^{-1}$ cloud in Figure\,\ref{fig:comp8_13co_494-03}(b) shows diffuse $^{13}$CO emissions around the footpoints of the arch-like structure. In addition, there are three CO components, which appear to surround the 21\,cm counters of G49.4-0.3, at $(l,b)\sim(49\fdg32, -0\fdg25)$, $(49\fdg33, -0\fdg36)$, and $(40\fdg46, -0\fdg31)$.
In Figure\,\ref{fig:comp8_13co_494-03}(c) the filamentary structures of the 60\,km\,s$^{-1}$ cloud shown in Figure\,\ref{fig:four_integ_13}(c) are plotted.
The 8\,$\mu$m filaments stretched nearly parallel to the Galactic plane are traced by the upper rim of the $^{13}$CO filamentary structure of the 60\,km\,s$^{-1}$ cloud at $l\sim49\fdg34$--$49\fdg43$ and $b\sim-0\fdg32$--$-0\fdg30$.
The $^{13}$CO filamentary structure harbors high-$R_{3210}^{13}$ gas at the same $l$ range, as shown in Figure\,\ref{fig:four_R3210}.
The 68\,km\,s$^{-1}$ cloud (HVS) in Figure\,\ref{fig:comp8_13co_494-03}(d) are distributed almost parallel to the filamentary structure in the 60\,km\,s$^{-1}$ cloud.

Figure\,\ref{fig:G49.4-0.3_pv}(a) presents the C$^{18}$O distributions of the four clouds toward G49.4-0.3 in the same manner as in Figure\,\ref{fig:G49.5-0.4_vb}(a).
While the 56\,km\,s$^{-1}$ cloud is not detected in C$^{18}$O, the 60 and 68\,km\,s$^{-1}$ clouds show fragmented distribution at $l\sim49\fdg34$--$49\fdg40$, and are aligned along the Galactic latitude with the 50\,km\,s$^{-1}$ cloud, showing a complementary distribution.
The $v$-$b$ diagram for this $l$ range is presented in Figure\,\ref{fig:G49.4-0.3_pv}(b), which shows that the 60 and 68\,km\,s$^{-1}$ clouds are bridged by the C$^{18}$O emissions at the intermediate velocities, while the 50 and 60\,km\,s$^{-1}$ clouds are connected with the $^{12}$CO emissions at $b\sim-0\fdg32$--$-0\fdg29$ in 55--60\,km\,s$^{-1}$, where the spatial distribution of the latter connecting feature is shown in Figure\,\ref{fig:comp8_13co_494-03}(b) in $^{13}$CO. 
These intermediate velocity features may be interpreted as the broad bridge features which suggest interactions between different velocity components.
Compared to the 8\,$\mu$m emissions in Figure\,\ref{fig:G49.4-0.3_pv}(a), the C$^{18}$O emissions of the 60 and 68\,km\,s$^{-1}$ clouds correspond to the regions \textcolor{black}{where the 8\,$\mu$m emission is faint. 
The column density $N({\rm H_2})$ of the 60\,km\,s$^{-1}$ and 68\,km\,s$^{-1}$ clouds are typically $0.6$--$1.0\, \times \, 10^{23}\ {\rm cm^{-2}}$ \textcolor{black}{(The column density maps are presented in Appendix\,\ref{app:cd_C18O})}, which corresponds to $A_{\rm V} \, \sim \,32$--$53$\,mag (\cite{2011MNRAS.412.1686S}). 
\citet{1989ApJ...345..245C} and \citet{2005ApJ...619..931I} reported $A_{\rm V}/A_{\rm K} \, \sim \, 8.8$ ($R_{\rm V} \, \sim \, 3.1$) and $A_{\rm [8.0\,\mu m]}/A_{\rm K} \, \sim \, 0.43$, respectively. 
Therefore, $N({\rm H_2})$ of $0.6$--$1.0 \, \times \, 10^{23}\ {\rm cm^{-2}}$ corresponds $A_{\rm [8.0\,\mu m]} \, \sim \, 1.5$--$2.6$\,mag. 
The faintness of the 8\,$\mu$m emission is considered to be an extinction by the 60\,km\,s$^{-1}$ and 68\,km\,s$^{-1}$ clouds, }suggesting these two clouds are both located in front of the nebulosities of G49.4-0.3 (\cite{2015A&A...573A.106G}), lending further support to the idea that these two clouds are distributed at the same location.
At the top of the arch-like structure of the 50\,km\,s$^{-1}$ cloud, where G49.4-0.3a and several MYSOs are distributed, we cannot find complementary distribution among different velocity components, while there is possibly a bridge feature in the $^{13}$CO emissions between the 50 and 60\,km\,s$^{-1}$ at $b\sim-0\fdg24$ (Figure\,\ref{fig:G49.4-0.3_pv}(b)).

\subsubsection{G49.57-0.27}
\begin{figure}[h]
 \begin{center}
   \includegraphics[width=7cm]{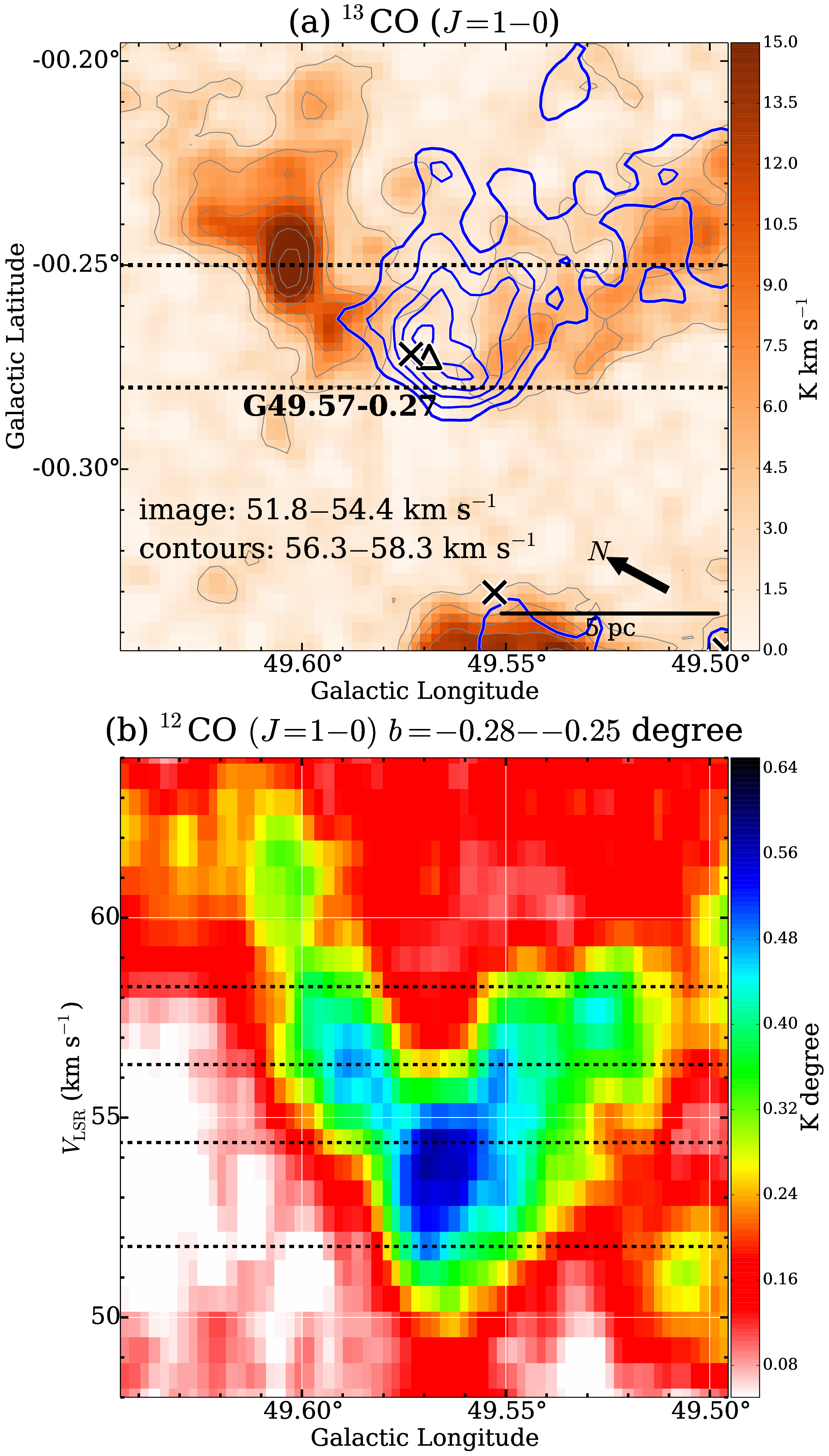}
 \end{center}
  \caption{(a) The complementary distributions of the two velocity components toward G49.57-0.27. The color scale shows $^{13}$CO ($J$=1--0) for 56.3--58.3\,km\,s$^{-1}$, while the blue contours show $^{13}$CO ($J$=1--0) for 51.8--54.4\,km\,s$^{-1}$. The blue contours are plotted with $3\,\sigma$ (4\,K\,km\,s$^{-1}$) intervals starting from the $5\,\sigma$ (6\,K\,km\,s$^{-1}$) level. Crosses represent compact H{\scriptsize II} regions listed in \citet{1994ApJS...91..713M} (Table\,\ref{tab:sources}), while the MYSO identified by \citet{2017ApJ...839..108S} is plotted with triangle. (b) Galactic Longitude-velocity diagram of the $^{12}$CO ($J$=1--0) emission toward G49.57-0.27. The integration range in $b$ is shown in (a) with vertical dashed lines. }\label{fig:G49.57-0.27_12CO}
\end{figure}

As presented in Figure\,\ref{fig:four_R3210}, G49.57-0.27 shows high $R_{3210}^{13}$ in the 50 and 56\,km\,s$^{-1}$ clouds.
G49.57-0.27 is an isolated compact H{\sc ii} region located at $(l,b)\sim(49\fdg57, -0\fdg27)$, whose ionizing photon flux measured from a 21\,cm continuum map corresponds to a spectral type of B0  (see Table\,\ref{tab:sources}).
Figure\,\ref{fig:G49.57-0.27_12CO}(a) shows the $^{13}$CO ($J$=1--0) integrated intensity maps of the 50 and 56\,km\,s$^{-1}$ clouds in contours and color, respectively. 
The CO emission in the 50\,km\,s$^{-1}$ cloud shows a roughly circular distribution with a diameter of $\sim$3\,pc, and is spatially coincident with G49.57-0.27 depicted by a cross in Figure\,\ref{fig:G49.57-0.27_12CO}(a). 
On the other hand, the 56\,km\,s$^{-1}$ cloud shows two components separated along the Galactic longitude, and 
the 50\,km\,s$^{-1}$ component is sandwiched by these two components, indicating a complementary distribution.
In the $l$-$v$ diagram in Figure\,\ref{fig:G49.57-0.27_12CO}(b), the 50\,km\,s$^{-1}$ component is connected with the two separated components in the 56\,km\,s$^{-1}$ cloud, showing a ``V-shape'' gas distribution in the $^{12}$CO ($J$=1--0) emissions.
As introduced in Section\,2, detections of V-shape gas distribution in the $p$-$v$ diagram were reported in several CCC regions (e.g.,  \cite{2018ApJ...859..166F, 2017arXiv170605652O, 2017arXiv170605871H, 2017arXiv170607164T}). 
Based on the synthetic CO observations performed by \citet{2015MNRAS.454.1634H} using the CCC \textcolor{black}{scenarios} by \citet{2014ApJ...792...63T}, \citet{2018ApJ...859..166F} and \citet{2017arXiv170607164T} reproduced the V-shape gas distribution in the $p$-$v$ diagram (see Figure\,14 of \citet{2018ApJ...859..166F}), which resembles the present CO observations shown in Figure\,\ref{fig:G49.57-0.27_12CO}(a).

\section{Discussion}\label{sec:Dis}

Our analyses of the new CO ($J$=1--0) data basically confirmed the observed features in the previous studies by \citet{1998AJ....116.1856C} and \citet{2001PASJ...53..793O}. In addition, our CO data including the C$^{18}$O ($J$=1--0) emission revealed previously unreported signatures that can be summarized as follows:
\begin{enumerate}
\item At the center of G49.5-0.4, in which IRS\,1 and IRS\,2 are located, the four C$^{18}$O clumps, which are each embedded within the four velocity clouds, show a complementary distribution within a small area less than 5\,pc \textcolor{black}{(Figure\,\ref{fig:C18O_comp})}. These are connected with each other in the $p$-$v$ diagram with $^{13}$CO and/or C$^{18}$O emissions \textcolor{black}{(Figure\,\ref{fig:G49.5-0.4_vb})}, suggesting broad bridge features.
\item In G49.4-0.3, the $^{13}$CO filamentary structures in the 50, 60, and 68\,km\,s$^{-1}$ clouds elongated nearly parallel to the Galactic plane are aligned \textcolor{black}{(Figure\,\ref{fig:four_integ_13})}. These filamentary structures show high $R_{3210}^{13}$ near G49.4-0.3 \textcolor{black}{(Figure\,\ref{fig:four_R3210})}, suggesting physical associations with G49.4-0.3. Each pair of the 50 and 60\,km\,s$^{-1}$ clouds and the 60 and 68\,km\,s$^{-1}$ clouds are connected with the bridge features in the $p$-$v$ diagram \textcolor{black}{(Figure\,\ref{fig:G49.4-0.3_pv})}.
\item In the relatively evolved H{\sc ii} regions with larger sizes, i.e., G49.5-0.4f, g, h, and i, we found no CO counterparts which are spatially coincident with the H{\sc ii} regions along the line of sight, but identified the remnant CO fragments in the 50, 56, and 60\,km\,s$^{-1}$ clouds \textcolor{black}{(e.g., Figure\,\ref{fig:four_integ_13})}, which have high $R_{3210}^{13}$ at the rims of these H{\sc ii} regions \textcolor{black}{(Figure\,\ref{fig:four_R3210})}.
\item In the isolated H{\sc ii} region G49.57-0.27, a complementary distribution between the gas components in the 50 and 56\,km\,s$^{-1}$ clouds are discovered \textcolor{black}{(Figure\,\ref{fig:G49.57-0.27_12CO})}, where the circular CO emission in the 50\,km\,s$^{-1}$ cloud is sandwiched by the two separated components in the 56\,km\,s$^{-1}$ cloud. The complementary distribution is seen as a V-shape gas distribution in the $l$-$v$ diagram, which is well reproduced by the numerical calculation of CCC.

\end{enumerate}
In this section, we discuss a CCC \textcolor{black}{scenario} in W51A based on the obtained results summarized above.

\subsection{Ages of the H{\sc ii} regions}\label{sec:age}
It is important to obtain the ages of the H{\sc ii} regions in W51A in order to discuss the formation mechanism of their exciting stars.
\citet{2000ApJ...543..799O} estimated the ages of the several H{\sc ii} regions listed in Table\,\ref{tab:sources} by measuring sizes of the H{\sc ii} regions. 
We calculated the ages of the remaining H{\sc ii} regions using the analytical model of the D-type expansion developed by \citet{1978ppim.book.....S}, where the sizes of the H{\sc ii} regions and classifications of the exciting sources summarized in Table\,\ref{tab:sources} were adopted. 
\textcolor{black}{This calculation method is the same as the method \citet{2000ApJ...543..799O} used.}
We assumed a uniform initial density of gas as $10^4$\,cm$^{-3}$, as the dense gas probed using C$^{18}$O is widely detected in the molecular clouds in W51A.
\textcolor{black}{The electron temperature was also assumed to be a constant value, 8000\,K \textcolor{black}{(e.g., \cite{1978ppim.book.....S})}. }
In Table\,\ref{tab:sources} the ages estimated in this study are marked by asterisks.
\textcolor{black}{
G49.5-0.4 includes H{\sc ii} regions with various ages (0.1--2.6\,Myr), while G49.4-0.3 includes only H{\sc ii} regions with ages of $<$1\,Myr. 
}

\subsection{CCC \textcolor{black}{scenarios} in W51A}\label{sec:CCCscenarios}

There are mainly two CCC \textcolor{black}{scenarios} in W51A discussed in the previous studies by \citet{1998AJ....116.1856C} and \citet{2001PASJ...53..793O}.
\citet{1998AJ....116.1856C} assumed that the present 50, 56 and 60\,km\,s$^{-1}$ clouds, which correspond to 53, 58, 60, and 63\,km\,s$^{-1}$ components in \citet{1998AJ....116.1856C}, are inner clouds of a single GMC (the W51 GMC). 
The authors discovered that the northern tip of the 68\,km\,s$^{-1}$ cloud is truncated at the location of the 60\,km\,s$^{-1}$ clouds.
Although no detailed process was discussed to create such a complementary distribution, \citet{1998AJ....116.1856C} postulated a CCC scenario between these the W51 GMC and the 68\,km\,s$^{-1}$ cloud.

On the other hand, based on the $^{13}$CO ($J$=1--0) observations in G49.5-0.4, \citet{2001PASJ...53..793O} discussed CCCs for two pairs of the clouds, i.e., the 56 and 60\,km\,s$^{-1}$ clouds and the 60 and 68\,km\,s$^{-1}$ clouds.
Although the authors found no plausible evidence of the collision between the 50 and 56\,km\,s$^{-1}$ clouds, they postulated a CCC \textcolor{black}{scenario} that four discrete molecular clouds distributed in a line along the line of sight are moving at different velocities, resulting in the ``pileup'' of these four clouds.

Our results provide new insight into the CCCs in the W51A region.
\textcolor{black}{Figures\,\ref{fig:C18O_comp} and }\ref{fig:G49.5-0.4_vb}(a) indicate that the 50, 56, 60, and 68\,km\,s$^{-1}$ clouds show a complementary distribution. 
As introduced in Section \ref{sec:obsCCC}, the recent works on CCC indicate that a complementary distribution can be created through a collision of two molecular clouds of different sizes or with a spatial offset (\cite{2011ApJ...738...46T, 2018ApJ...859..166F}). 
If so, the present results suggest that multiple collisions of the four clouds have \textcolor{black}{perhaps} occurred in G49.5-0.4, resulting in the formation of the massive stars at the interfaces of the collisions.
In this scenario, it is reasonably assumed that the observed C$^{18}$O clumps were formed through strong compression by the collisions. 
That the several compact H{\sc ii} regions are concentrated around the interfaces of the complementary distribution lends more credence to this CCC scenario. 
\textcolor{black}{
In Figure\,\ref{fig:C18O_comp}(a), we could not observe a clear complementary distribution like Figures\,\ref{fig:C18O_comp}(b) and \ref{fig:C18O_comp}(c).
}

\textcolor{black}{
In G49.5-0.4, the total mass of the dense gas is estimated to be $\sim 5\,\times \,10^4$\,M$_{\odot}$, and $\sim \,$30 O-stars have been identified (\cite{2000ApJ...543..799O}).
If we assume that (1) one O-star was formed in one massive core and (2) the mass of each massive core is 100\,M$_{\odot}$, which is a value used as an initial mass of the massive core in the simulation of the massive star formation performed by \citet{2009Sci...323..754K}, the massive core formation efficiency in the dense gas is estimated to be $\sim \,$6\%.
This estimation is coarse.
To estimate with high accuracy, observational studies with higher spatial resolution are required.
}

The timescale of the collisions \textcolor{black}{in G49.5-0.4} can be approximately estimated from a ratio of the cloud size and the relative velocity between the two clouds. 
If we assume the sizes of the four clumps of $\sim$2\,pc (Table\,\ref{tab:C18O}) and the relative velocities of 4--18\,km\,s$^{-1}$, the estimated timescales of the collisions in G49.5-0.4 ranges 0.1--0.5\,Myrs.
These figures are consistent with the estimated ages of the H{\sc ii} regions distributed around the interfaces of the complementary distribution, which include IRS\,1 and IRS\,2 (Table\,\ref{tab:sources}).
While our results indicate that the 50, 56, and 60\,km\,s$^{-1}$ clouds are almost blended within a small volume, which is consistent with the discussion by \citet{2015A&A...573A.106G} based on the H$_2$CO absorption observations, spatial correlation of the 68\,km\,s$^{-1}$ cloud with the extinction in the 8\,$\mu$m emission (see Figure\,\ref{fig:comp8_c18o}(d)) shows that it is located in front of G49.5-0.4, not blended with the other three clouds.


In G49.4-0.3, the complementary distribution of gas and presence of the bridge features suggest collisions of the 50 and 60\,km\,s$^{-1}$ clouds and the 60 and 68\,km\,s$^{-1}$ clouds (Figure\,\ref{fig:G49.4-0.3_pv}).
As the filamentary structures in the 60\,km\,s$^{-1}$ cloud appear to surround the 21\,cm continuum emissions of G49.4-0.3 (Figure\,\ref{fig:four_integ_13}(c)), it is also possible to interpret the associations of the multiple velocity components in G49.4-0.3 as being to expansion of the H{\sc ii} regions.
If so, as the 60\,km\,s$^{-1}$ cloud is redshifted relative to the systemic velocity of the H{\sc ii} regions in G49.4-0.3 having velocities of $\sim$52\,km\,s$^{-1}$ \citep{2015A&A...573A.106G}, it should be located behind G49.4-0.3. However, as seen in Figure\,\ref{fig:G49.4-0.3_pv}(a), the C$^{18}$O components in the 60\,km\,s$^{-1}$ cloud coincide with the extinction in the 8\,$\mu$m image. 
This indicates that it is located in front of G49.4-0.3 \citep{2015A&A...573A.106G}, against the assumption of the H{\sc ii} region expansion.
On the other hand, in the CCC scenario, the bridge feature in the $p$-$v$ diagram indicates that the collision is on-going, and it is possible that the 60\,km\,s$^{-1}$ cloud has not completely passed over G49.4-0.3 along the line of sight yet.
In the CCC scenario, the dissipative effect of the H{\sc ii} regions still can work on the 60\,km\,s$^{-1}$ cloud to form CO components which surround G49.4-0.3 (\textcolor{black}{indicated by green arrows in Figure\,\ref{fig:four_integ_13}(c)}), suggesting that the 60\,km\,s$^{-1}$ cloud originally had an extended gas distribution in this region.

The timescale of the collision between the 50 and 60\,km\,s$^{-1}$ cloud in G49.4-0.3 can be estimated as 0.2--0.6\,Myr by assuming collision length ranging from the width of the filamentary structure, $\sim$2\,pc, to the full extension of the H{\sc ii} regions in G49.4-0.3, $\sim$6\,pc.
On the other hand, as we found no plausible evidence of physical association between the 68\,km\,s$^{-1}$ cloud and the H{\sc ii} regions in G49.4-0.3, it may be that the 60 and 68\,km\,s$^{-1}$ clouds are in the beginning of the collision, and compressed dense gas is observed in C$^{18}$O as shown in Figure\,\ref{fig:G49.4-0.3_pv}(a).

In G49.57-0.27, the V-shape gas distribution in the $l$-$v$ diagram in Figure\,\ref{fig:G49.57-0.27_12CO}(b) may also be interpreted by expansion of the H{\sc ii} region.
However, as the radius of G49.57-0.27, $\sim$1.2\,pc (Table\,\ref{tab:sources}), is much smaller than that of the 50\,km\,s$^{-1}$ component with 3--4\,pc, the observed V-shape gas distribution cannot be attributed to expansion of the H{\sc ii} region. It is therefore more likely that the V-shape was formed through the CCC process. 
\textcolor{black}{If we assume a collision angle relative to the line of sight $\theta = 45^{\circ}$,} the timescale required for the 50\,km\,s$^{-1}$ cloud to completely punch the 56\,km\,s$^{-1}$ cloud can be estimated as 3--4\,pc / 6\,km\,s$^{-1}$\,$\sim$\,0.5--0.7\,Myr.
\textcolor{black}{Although we could not determine an accurate value of $\theta$ from position-position-velocity data of molecular clouds, $\theta = 45^{\circ}$ is one of the convenient solutions as a first approximation of the angle of two colliding clouds (the error of $<$\,5 times is within the range of 86\%). }

\textcolor{black}{
Meanwhile, it is difficult to identify a CCC event in position-position-velocity data when the collision angle is parallel to the sky plane, and thus, there may be unidentified CCCs in W51A. 
\textcolor{black}{
A fraction of such unidentified CCC events with collide angle of $>\,80^{\circ}$ is estimated to be less than $20$\%, if we tentatively assume that the cloud motions are all random (See Appendix\,\ref{app:CCC_prob}).}
To discuss these unidentified CCC events, we need to conduct statistical study of CCCs in the Galaxy.
}

\textcolor{black}{\subsection{Massive star formation triggered by CCCs in W51A}}\label{sec:MSF}

G49.5-0.4 \textcolor{black}{also} includes relatively evolved H{\sc ii} regions with ages of a few Myrs, i.e., G49.5-0.4f, g, h, and i (Table\,\ref{tab:sources}). 
As summarized in the beginning of \ref{sec:CCCscenarios}, the remnant CO components of these H{\sc ii} regions are found in the 50, 56, and 60\,km\,s$^{-1}$ clouds. 
These observed signatures can be interpreted as a CCC scenario or expansion of the H{\sc ii} regions. 
In the evolved H{\sc ii} regions, complementary distributions and bridge features are dissipated and cannot be observed. 
Therefore, \textcolor{black}{in G49.5-0.4f, g, h, and i}, it is difficult to conclude whether the exciting massive stars in these H{\sc ii} regions were formed via CCCs or not.
If we tentatively assume that these H{\sc ii} regions were also formed via the collisions among these three clouds, the collisions should have occurred since a few Myrs ago.
The sequential collisions and star formation from the north to the south are consistent with the discussion by \citet{2001PASJ...53..793O}. 

\textcolor{black}{In G49.4-0.3,} the derived figures \textcolor{black}{in section \ref{sec:CCCscenarios}} are consistent with the estimated ages of the H{\sc ii} regions \textcolor{black}{(Table \ref{tab:sources})}, \textcolor{black}{implying a scenario} that the collision triggered formation of massive stars. 
\textcolor{black}{In G49.57-0.27,} the collision timescale of 0.5--0.7\,Myr \textcolor{black}{derived in section \ref{sec:CCCscenarios}} is \textcolor{black}{also consistent with} the estimated age of the H{\sc ii} region of G49.57-0.27, \textcolor{black}{0.7\,Myr} (Table\,\ref{tab:sources}).
As the broad bridge feature indicates that the collision is still continuing, the collision in G49.57-0.27 is likely in the middle, which is consistent with the young age of G49.57-0.27. 

\begin{figure}[h]
 \begin{center}
   \includegraphics[width=17cm]{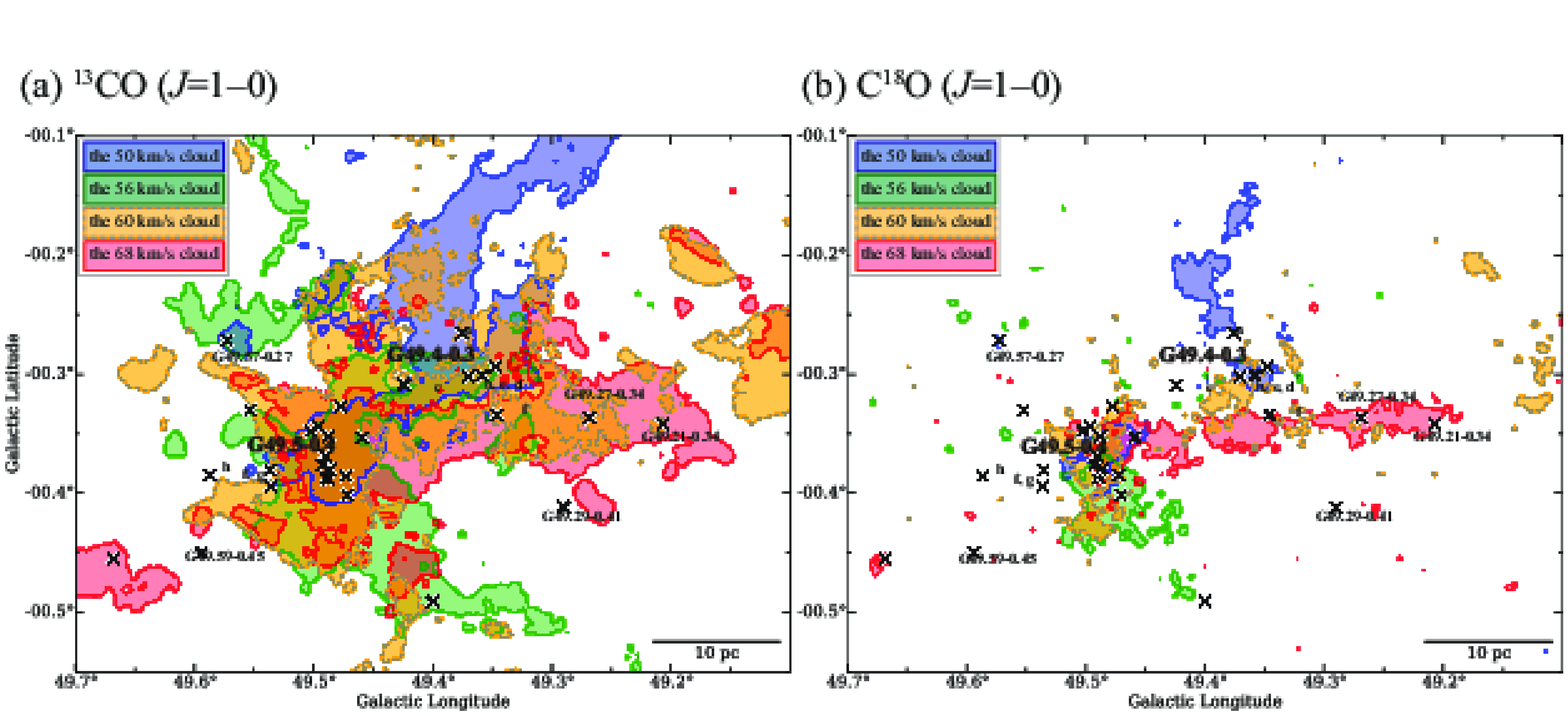}
 \end{center}
  \caption{\textcolor{black}{(a) The $^{13}$CO ($J$=1--0) integrated intensity distributions of the 50 (blue contour), 56 (green contour), 60 (orange contour with gray dashed-contour), and 68\,km\,s$^{-1}$ clouds (red contour). The integration ranges is the same as Figure \ref{fig:four_integ_13}, and the contour level is 10, 10, 10, and 8\,K\,km\,s$^{-1}$, respectively. The crosses represent H{\sc ii} regions listed in \citet{1994ApJS...91..713M}. (b) Same as (a), but for the C$^{18}$O ($J$=1--0) integrated intensity distributions. }}\label{fig:sketch1}
\end{figure}

\begin{figure}[h]
 \begin{center}
   \includegraphics[width=17cm]{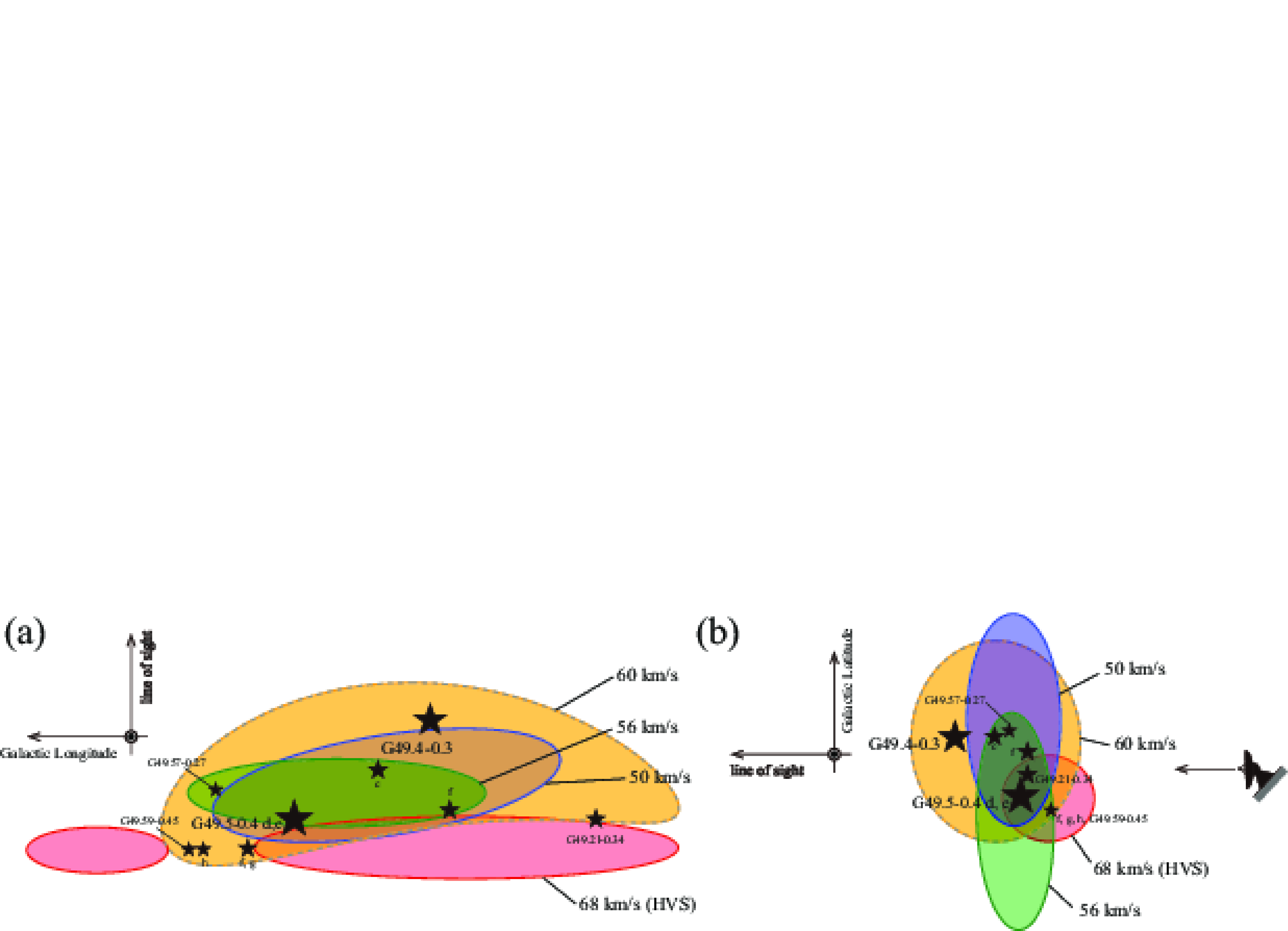}
 \end{center}
  \caption{\textcolor{black}{(a) The schematic pictures of the distributions of the molecular gas in W51A based on \citet{2015A&A...573A.106G} as viewed from the Galactic north pole. Star markers indicate the positions of the representative H{\sc ii} regions in W51A. (b) Same as (a), but as viewed from the Galactic eastern side. }}\label{fig:sketch2}
\end{figure}

Summarizing the discussions \textcolor{black}{in Subsection \ref{sec:age}--\ref{sec:CCCscenarios}}, the observational signatures of CCCs in G49.5-0.4, G49.4-0.3 and G49.57-0.27 represent the collisions which have started since a few 0.1\,Myr ago.
This indicates that the four velocity components in this region, i.e., the 50, 56, 60, and 68\,km\,s$^{-1}$ clouds, are currently distributed close to each other, and are partly blended into one single molecular cloud \textcolor{black}{as shown in Figures\,\ref{fig:sketch1}(a) and \ref{fig:sketch1}(b), which show the $^{13}$CO ($J$=1--0) and C$^{18}$O ($J$=1--0), respectively, of the four clouds. }
\textcolor{black}{Besides, Figures\,\ref{fig:sketch2}(c) and \ref{fig:sketch2}(d) show the schematic pictures of the $^{13}$CO ($J$=1--0) distributions in W51A based on \citet{2015A&A...573A.106G} as viewed from the Galactic north pole and the Galactic eastern side, respectively. }

This is against for the assumption by \citet{2001PASJ...53..793O} that the four clouds are located in a line along the line of sight. 
However, in the latter evolutionary stage of their ``pileup'' scenario, it can be expected that the four clouds are completely merged into a single cloud (see Figure\,5 of \cite{2001PASJ...53..793O}). 
In this sense, our results are consistent with their CCC scenario.

\begin{figure}[h]
 \begin{center}
   \includegraphics[width=17cm]{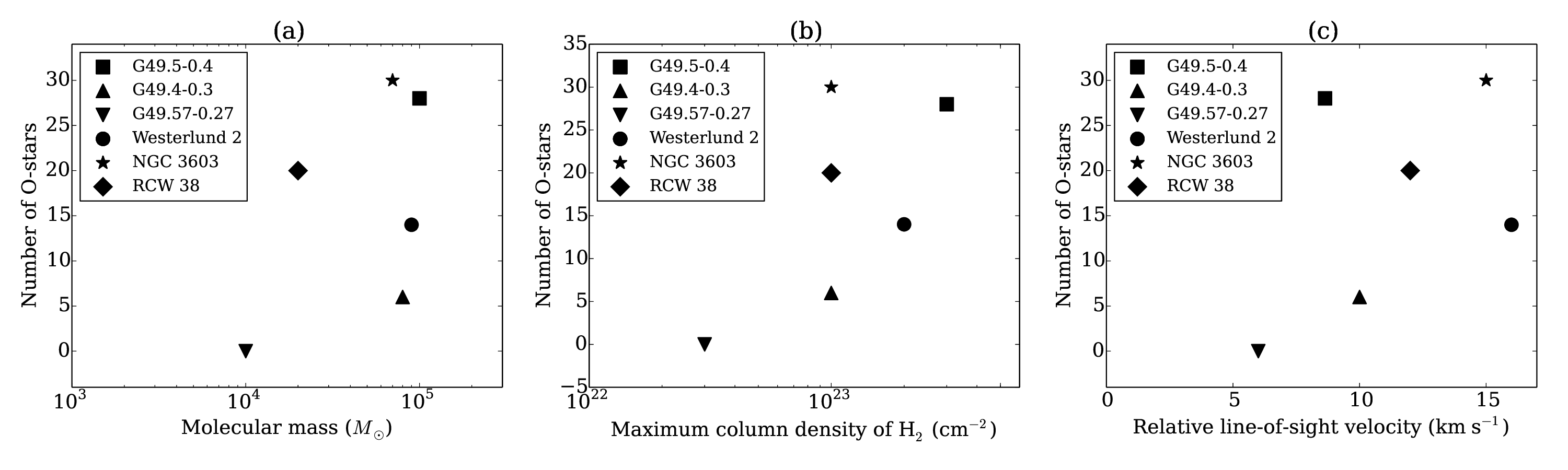}
 \end{center}
  \caption{\textcolor{black}{Correlation between (a) the molecular mass and the number of O-stars, (b) the $N_{\rm max}$(H$_2$) and the number of O-stars, and (c) the relative line-of-sight velocity and the number of O-stars. In (a) and (b), the molecular masses and the $N_{\rm max}$(H$_2$) of the largest cloud in the each H{\sc ii} region are plotted. In (c), the geometric mean of the relative velocities is plotted for G49.5-0.4. }}\label{fig:sca_col}
\end{figure}

\textcolor{black}{
In Table\,\ref{tab:CCC}, we compare the properties of the colliding molecular clouds in W51A with those of the other H{\sc ii} regions containing more than 10 O-stars formed by CCC discussed in previous studies. 
The number of O-stars in G49.5-0.4 and G49.4-0.3 are roughly comparable with those in Westerlund 2, NGC 3603, and RCW 38.
Furthermore, the H$_2$ column density of each the larger cloud is also roughly comparable ($1$\,--\,$3\,\times 10^{23}$\,cm$^{-2}$).
However, the molecular mass of the larger cloud in RCW 38 is significantly smaller than that of the other regions.
The number of O-stars formed by CCC possibly depend on H$_2$ column densities rather than molecular masses. 
The molecular clouds in G49.57-0.27 could not form O-stars because the H$_2$ column density is low.
On the other hand, we can not discuss the relationship between relative \textcolor{black}{line-of-sight velocity} separations and the number of O-stars from these data because of a large ambiguity of colliding velocity in 3-dimensional velocity.
\textcolor{black}{Figures\,\ref{fig:sca_col}(a)--(c) show the correlations between the number of O-stars and the parameters of the molecular cloud in the H{\sc ii} regions listed in Table\,\ref{tab:CCC}.
As mentioned above, the correlation between the $N_{\rm max}$(H$_2$) and the number of O-stars (Figure\,\ref{fig:sca_col}(b)) is likely stronger than that of between the molecular mass and the number of O-stars (Figure\,\ref{fig:sca_col}(a)). 
In addition, there may be a stronger positive correlation also between the number of O-stars and the relative line-of-sight velocity (Figure\,\ref{fig:sca_col}(c)). }
To establish a quantitative scenario for forming massive stars via a CCC, more studies, such as statistical studies and simulational studies, are required.
}

 \begin{table}[h]
   \tbl{\textcolor{black}{Physical properties of molecular clouds toward H{\sc ii} regions formed by cloud--cloud collisions}}
   {%
    \begin{tabular}{ccccccc}
    \hline \hline
    Name & Molecular masses & $N_{\rm max}({\rm H_2})$ & Relative \textcolor{black}{line-of-sight velocities} & Age & Number of O-stars & References \\
         & ($M_{\odot}$) & (cm$^{-2}$) & (km\,s$^{-1}$) & (Myr) & & \\  
    (1) & (2) & (3) & (4) & (5) & (6) & (7) \\   
       \hline \hline
   G49.5-0.4 & $\sim \,4\,\times \, 10^4$ &  $9\,\times \, 10^{22}$ & $6$\,--\,$20$ & $\sim \,0.1$\,--\,$2.6$ & $28$ & This study \& [1] \\
       ... & $\sim \,1\,\times \, 10^5$ &  $3\,\times \, 10^{23}$ & & & & \\ 
       ... & $\sim \,1\,\times \, 10^5$ &  $2\,\times \, 10^{23}$ & & & & \\ 
       ... & $\sim \,1\,\times \, 10^5$ &  $7\,\times \, 10^{22}$ & & & & \\
   G49.4-0.3 & $\sim \,8\,\times \, 10^4$ &  $1\,\times \, 10^{23}$& $10$ & $\sim \,0.1$\,--\,$0.8$ & $\sim \, 6$ & This study \& [1] \\
       ... & $\sim \,9\,\times \, 10^4$ &  $4\,\times \, 10^{22}$& & & & \\
   G49.57-0.27 & $\sim \,1\,\times \, 10^4$ &  $3\,\times \, 10^{22}$& $6$ & $\sim \,0.7$ & $0$ (one B0-star)& This study \& [1] \\
       ... & $\sim \,4\,\times \, 10^3$ &  $3\,\times \, 10^{22}$& & & & \\ \hline
   Westerlund 2 & $9\,\times \, 10^4$ & $2\,\times \, 10^{23}$ & 16 & $\sim \, 2.0$ &  $14$& [2], [3] \\ 
       ... & $8\,\times \, 10^4$ & $2\,\times \, 10^{22}$ &   &    &    & \\
   NGC 3603 & $7\,\times \, 10^4$ & $1\,\times \, 10^{23}$ & 15 & $\sim \, 2.0$ &  $\sim \, 30$ & [4] \\ 
       ... & $1\,\times \, 10^4$ & $1\,\times \, 10^{22}$ &    &    &    & \\
   RCW 38 & $2\,\times \, 10^4$ & $1\,\times \, 10^{23}$ & 12 & $\sim \, 0.1$ &  $\sim \, 20$ & [5] \\ 
         ... & $3\,\times \, 10^3$ & $1\,\times \, 10^{22}$ &    &    &    & \\
       \hline
    \end{tabular}} \label{tab:CCC}
\begin{tabnote}
(1) Name of H{\sc ii} regions. 
(2) Molecular mass derived from the $^{12}$CO ($J$=1--0) and $^{13}$CO ($J$=1--0)  data by assuming LTE.  
(3) Maximum H$_2$ column density derived from the $^{12}$CO ($J$=1--0) and $^{13}$CO ($J$=1--0)  data by assuming LTE.
(4) Relative \textcolor{black}{line-of-sight velocity} separation among the colliding clouds. 
(5) Age of H{\sc ii} regions. 
(6) Number of O-stars. 
(7) References. [1] \citet{2000ApJ...543..799O}, [2] \citet{2009ApJ...696L.115F}, [3] \citet{2010ApJ...709..975O}, [4] \citet{2014ApJ...780...36F}, and [5] \citet{2016ApJ...820...26F}.
\end{tabnote}
\end{table}

\section{Summary}
We carried out new $^{12}$CO ($J$=1--0), $^{13}$CO ($J$=1--0), and C$^{18}$O ($J$=1--0) observations toward W51A as a part of the FUGIN project with the Nobeyama 45-m telescope. 
These observations covered a large area of W51A ($1\fdg 4 \times 1\fdg 0$) at an angular resolution of 20$''$ ($\sim0.5$\,pc). 
The main conclusions of the present study are summarized as follows:

\begin{enumerate}
\item Our CO data identified four discrete velocity clouds with sizes and masses of $\sim 30$\,pc and $1.0$--$1.9\times10^5\ M_{\odot}$ at radial velocities of 50, 56, 60, and 68 km s$^{-1}$ in W51A. These four clouds mainly consist of the bright CO emissions toward the two bright H{\sc ii} region complexes G49.5-0.4 and G49.4-0.3 attached with the \textcolor{black}{filament hub} structures elongated for several tens pc.
\item Based on comparisons between our $^{13}$CO ($J$=1--0) data and the archival the JCMT $^{13}$CO ($J$=3--2) data, it was revealed that all of these four clouds are physically associated with G49.5-0.4, while three of the four, i.e., 50, 60, and 68\,km\,s$^{-1}$ clouds, are interact with G49.4-0.3, as the $^{13}$CO ($J$=3--2)/$^{13}$CO ($J$=1--0) intensity ratios in these clouds are increased higher than $1.0$ near the H{\sc ii} regions. The LVG calculations indicate such high ratios can be attributed to high temperature gas heated by the massive stars in these regions. We also found that the isolated H{\sc ii} region G49.57-0.27 located $\sim15$\,pc north of G49.5-0.4 are associated with the 50 and 56\,km\,s$^{-1}$ clouds.
\item In each of these three H{\sc ii} regions G49.4-0.5, G49.4-0.3, and G49.25-0.27, we revealed that the multiple velocity components associated with the H{\sc ii} regions show ``spatially complementary distributions'' on the sky and ``broad bridge features'' in the position-velocity diagrams. In particular, in G49.25-0.27 a combination of the complementary distribution and the bridge features represent a ``V-shape'' gas distribution in the position-velocity diagram. These signatures have been discussed as the observational signatures of CCC in recent theoretical and observational studies in the galactic H{\sc ii} regions. 
\item We estimated the timescales of the collisions in these three regions to be several 0.1\,Myrs by calculating crossing times of the collisions. These estimates are consistent with the ages of the H{\sc ii} regions measured from the sizes of the H{\sc ii} regions with the 21\,cm continuum map.
\item Our present results lend more credence to the CCC scenario in W51A, that multiple velocity components have been continuously colliding with each others, resulting in active massive star formation in W51A. 
\item \textcolor{black}{On the other hand, our results are also consistent with the discussion by \citet{1998AJ....116.1856C} that the 50, 56, and 60\,km\,s$^{-1}$ clouds represent kinematic structure within a single molecular cloud, as the total molecular mass in W51 is consistent with its virial mass for 100\,pc scale, suggesting that self-gravity will play a critical role in the evolution of the molecular clouds in W51. To fully understand the kinematics and interactions of molecular clouds in W51, which will allow us to investigate the future of this region, it is important to study W51B using the same CO dataset. We will work on this issue in a separate paper.}
\end{enumerate}

\clearpage

\begin{ack}
This study was financially supported by Grants-in-Aid for Scientific Research (KAKENHI) of the Japanese society for the Promotion of Science (JSPS; grant numbers 15K17607, 17H06740, and 18K13580). 
The authors would like to thank the all members of the 45-m group of Nobeyama Radio Observatory for support during the observation. 
Data analysis was carried out on the open use data analysis computer system at the Astronomy Data Center (ADC), of the National Astronomical Observatory of Japan (NAOJ), and made use of Astropy, a community-developed core Python package for Astronomy (\cite{2013A&A...558A..33A}),  APLpy, an open-source plotting package for Python (\cite{2012ascl.soft08017R}), astrodendro, a Python package to compute dendrograms of Astronomical data (\cite{2008ApJ...679.1338R}), and SCIMES, a Python package to find relevant structures into dendrograms of molecular gas emission using the spectral clustering approach (\cite{2015MNRAS.454.2067C}).
The authors also would like to thank NASA, National Radio Astronomy Observatory (NRAO), and Dr. H. Parsons for providing FITS data of {\it Spitzer} Space Telescope and the JCMT, and VGPS, respectively. 
\end{ack}

\clearpage

\appendix
\section{Velocity channel maps of the CO ($J$=1--0) emissions}\label{app:chmap}
In Figures\,\ref{fig:chmap}(a)--(c), we presented velocity channel maps of the $^{12}$CO ($J$=1--0), $^{13}$CO ($J$=1--0) and C$^{18}$O ($J$=1--0) emissions at a velocity step of 3.25\,km\,s$^{-1}$.

\begin{figure*}[h]
 \begin{center}
  \includegraphics[width=21cm, angle=90]{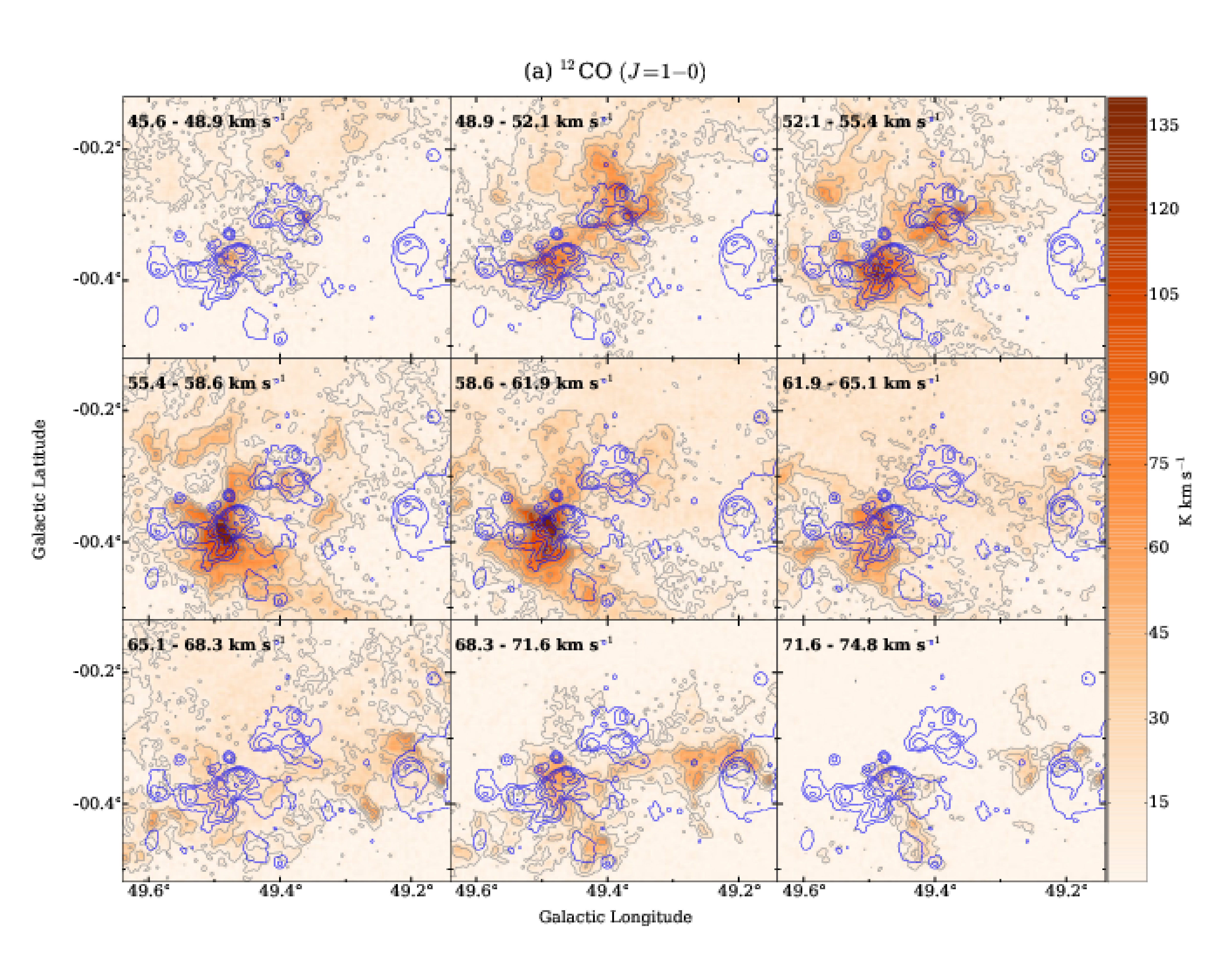}
 \end{center}
 \caption{The velocity channel maps of the $^{12}$CO ($J$=1--0) emissions. The gray contours are plotted with $8\,\sigma$ (16\,K\,km\,s$^{-1}$) intervals starting from the $4\,\sigma$ (8\,K\,km\,s$^{-1}$) level. Integration range in each panel is presented in the top-left corner of the panel. The blue contours indicate \textcolor{black}{the THOR 21\,cm radio continuum emission combined with the VGPS data (\cite{2016A&A...595A..32B, 2006AJ....132.1158S}), and are plotted from 0.03 (dashed lines) to 3.0\,Jy\,str$^{-1}$ with logarithm step. The angular resolution of the THOR data combined with the VGPS is 25$''$. } }\label{fig:chmap}
\end{figure*}

\begin{figure*}[h]
 \begin{center}
  \includegraphics[width=21cm, angle=90]{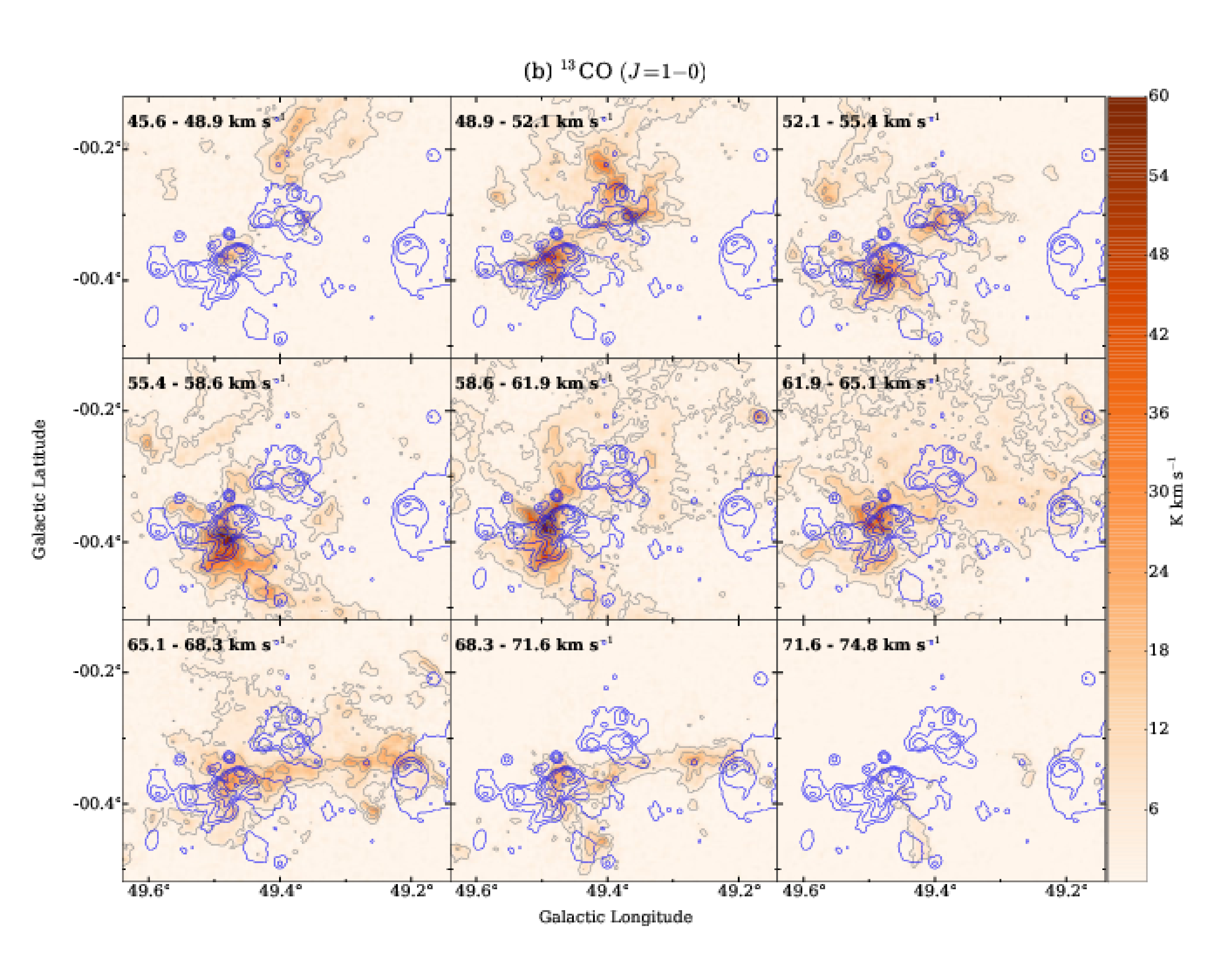}
 \end{center}
  \contcaption{(Continued. ) Same as \ref{fig:chmap} but for $^{13}$CO ($J$=1--0). The gray contours are plotted with $8\,\sigma$ (8\,K\,km\,s$^{-1}$) intervals starting from the $4\,\sigma$ (4\,K\,km\,s$^{-1}$) level.  }
\end{figure*}

\begin{figure*}[h]
 \begin{center}
  \includegraphics[width=21cm, angle=90]{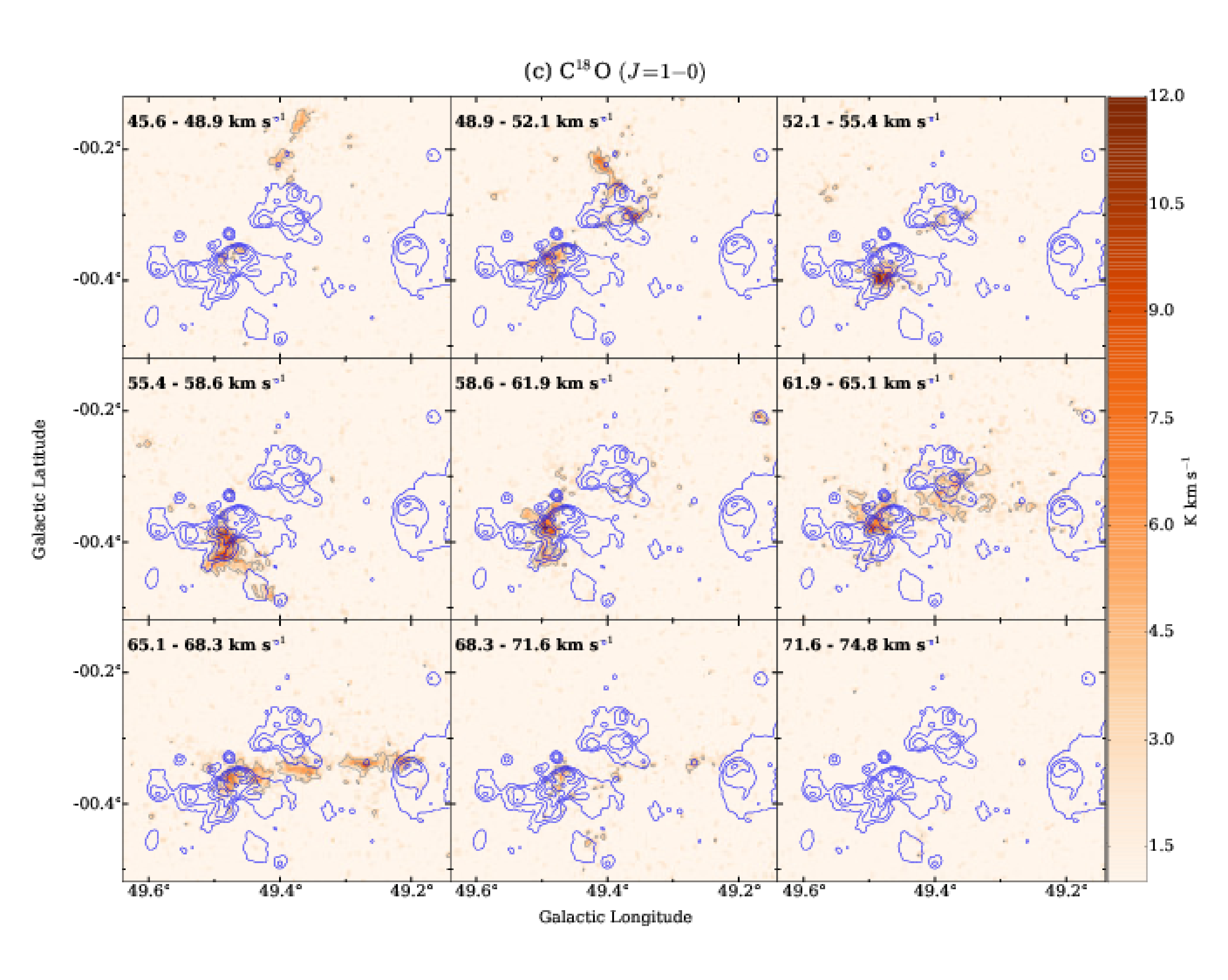}
 \end{center}
  \contcaption{(Continued. ) Same as \ref{fig:chmap} but for C$^{18}$O ($J$=1--0). The gray contours are plotted with $3\,\sigma$ (3\,K\,km\,s$^{-1}$) intervals starting from the $3\,\sigma$ (3\,K\,km\,s$^{-1}$) level. }
\end{figure*}

\clearpage
\section{\textcolor{black}{Errors associated with $R_{3210}^{13}$}}\label{app:ratio_error}

\begin{figure}[htbp]
  \begin{center}
         \includegraphics[width=16cm]{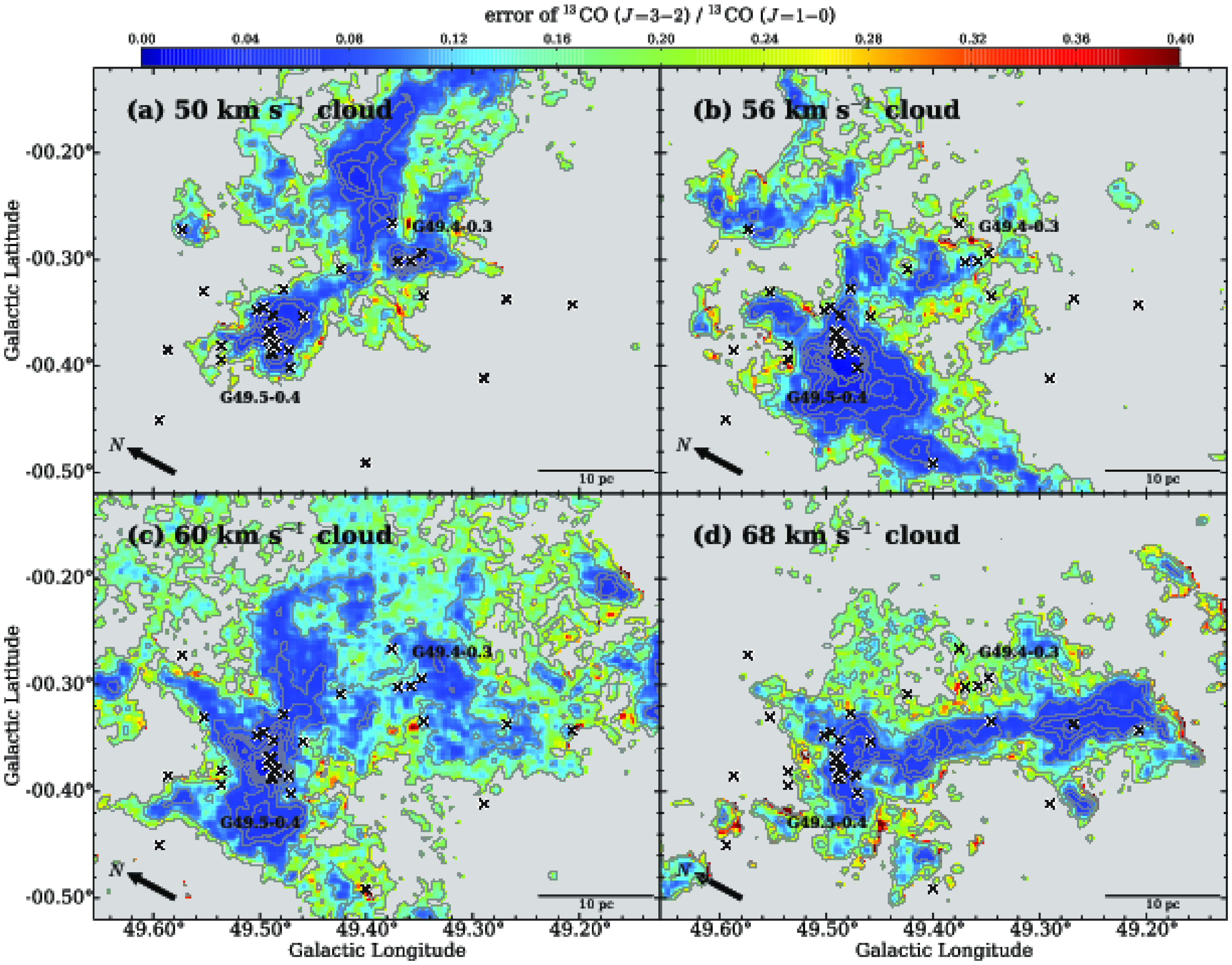}
  \end{center}
  \caption{\textcolor{black}{The map of the error associated with $R_{3210}^{13}$ (Figure\,\ref{fig:four_R3210}(a)--(d)) for each pixels. The errors were estimated from the calibration error of the $^{13}$CO ($J$=1--0) and $^{13}$CO ($J$=3--2) data (15\% and 10\%, respectively). }}\label{fig:ratio_error}
\end{figure}

\clearpage
\section{Large velocity gradient analysis}\label{app:LVG}
To investigate high-temperature gas in the molecular clouds of W51A, we utilize the large velocity gradient (LVG) calculations (e.g., \cite{1974ApJ...189..441G}).
The assumption of the uniform velocity gradient is not always valid in the molecular gas associated with H{\sc ii} regions. However radiative transfer calculations assuming a micro-turbulent cloud interacting with an H{\sc ii} region shows no significant difference from the LVG analysis (e.g., \cite{1976ApJ...208..732L, 1977ApJ...211..744W}). We therefore adopt the LVG approximation in the present study.
We here adopted the abundance ratios of [$^{12}$CO]/[$^{13}$CO] = 77 (\cite{1994ARA&A..32..191W}) and the fractional CO abundance to be X(CO) = [$^{12}$CO]/[H$_2$]\,=\,$10^{-4}$ (e.g., \cite{1982ApJ...262..590F, 1984ApJS...56..231L}). 
Two velocity gradient $dv/dr$ of 5 and 10\,km\,s$^{-1}$\,pc$^{-1}$ were adopted, by assuming a typical velocity width of individual velocity clouds, $\sim$3\,km\,s$^{-1}$ and a full velocity width of the four velocity clouds, $\sim$30\,km\,s$^{-1}$ (see Figure\,\ref{fig:lvs}), multiplied by a typical size of the molecular gas components of $\sim$3\,pc.

\begin{figure}[htbp]
  \begin{center}
         \includegraphics[width=8cm]{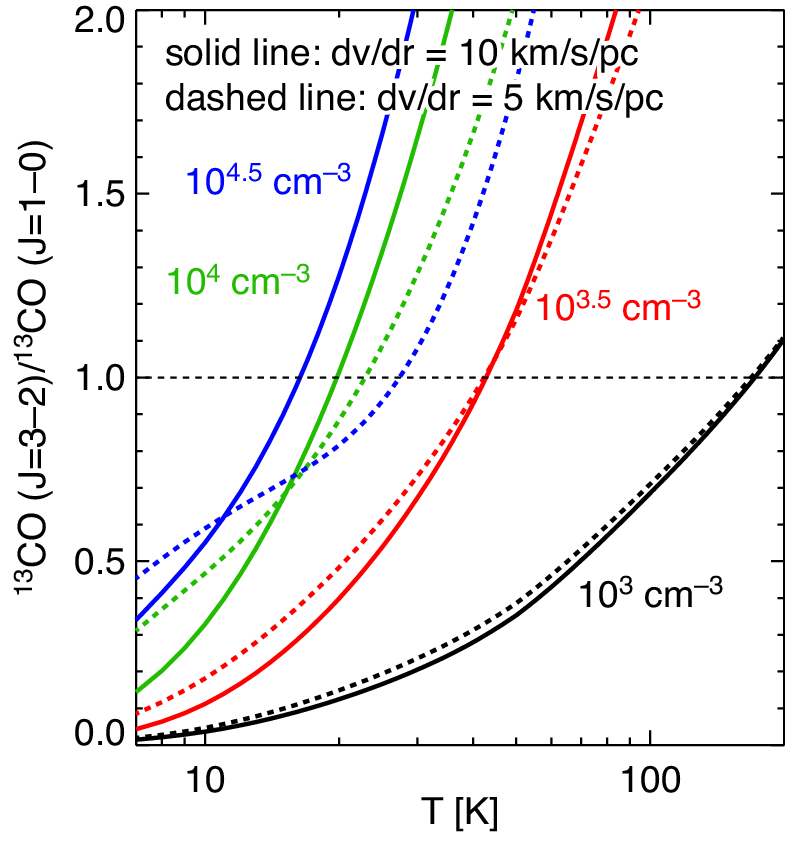}
  \end{center}
  \caption{Curves of $R_{3210}^{13}$ as a function of $T_{\rm k}$ and $n({\rm H_2})$, estimated using the LVG calculations. $dv/dr$ is assumed 10\,km\,s$^{-1}$\,pc$^{-1}$ (solid lines) and 5\,km\,s$^{-1}$\,pc$^{-1}$ (dashed lines). }\label{fig:LVG}
\end{figure}

Figure\,\ref{fig:LVG} shows $R_{3210}^{13}$ distributions calculated with LVG for different densities $n({\rm H_2})$ as a function of kinetic temperature $T_{\rm k}$ of gas, indicating that $R_{3210}^{13}$ is sensitive to gas temperature.  In all the $n({\rm H_2})$ cases, $R_{3210}^{13}$ of higher than 1.0 corresponds to $T_{\rm k}$ of higher than $\sim$20\,K.  20\,K is significantly higher than typical temperature of molecular clouds without star formation, 10\,K \citep{2016ApJ...820...26F}.

\clearpage
\section{\textcolor{black}{H$_2$ column density derived from the $^{12}$CO ($J$=1--0) and C$^{18}$O ($J$=1--0) emission}}\label{app:cd_C18O}
\begin{figure}[htbp]
  \begin{center}
         \includegraphics[width=16cm]{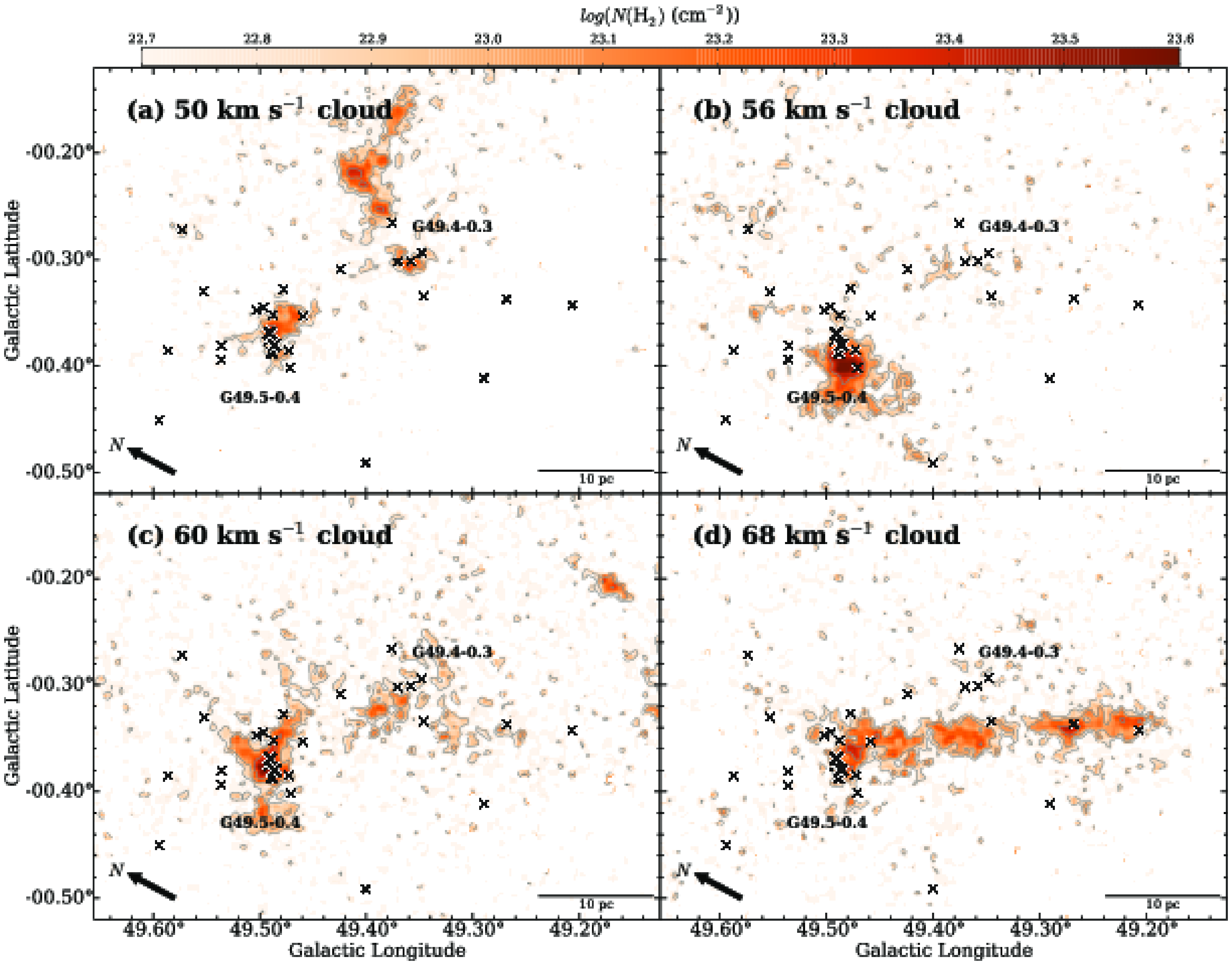}
  \end{center}
  \caption{\textcolor{black}{The column density of H$_2$ molecules (derived from the $^{12}$CO ($J$=1--0) and C$^{18}$O ($J$=1--0) emission) of the (a) 50, (b) 56, (c) 60, and (d) 68\,km\,s$^{-1}$ clouds, with the integration ranges of 46.9--52.1, 52.8--58.6, 59.3--64.5, and 65.1--71.0\,km\,s$^{-1}$, respectively. The gray contours are plotted with $6\, \times \, 10^{22}$\,cm$^{-2}$, $1\, \times \, 10^{23}$\,cm$^{-2}$, and $2\, \times \, 10^{23}$\,cm$^{-2}$ level. The crosses represent H{\sc ii} regions listed in \citet{1994ApJS...91..713M}.}}\label{fig:cd_C18O}
\end{figure}

\clearpage
\section{\textcolor{black}{Fraction of the collision angle with random motion}}\label{app:CCC_prob}
\begin{figure}[htbp]
  \begin{center}
         \includegraphics[width=8cm]{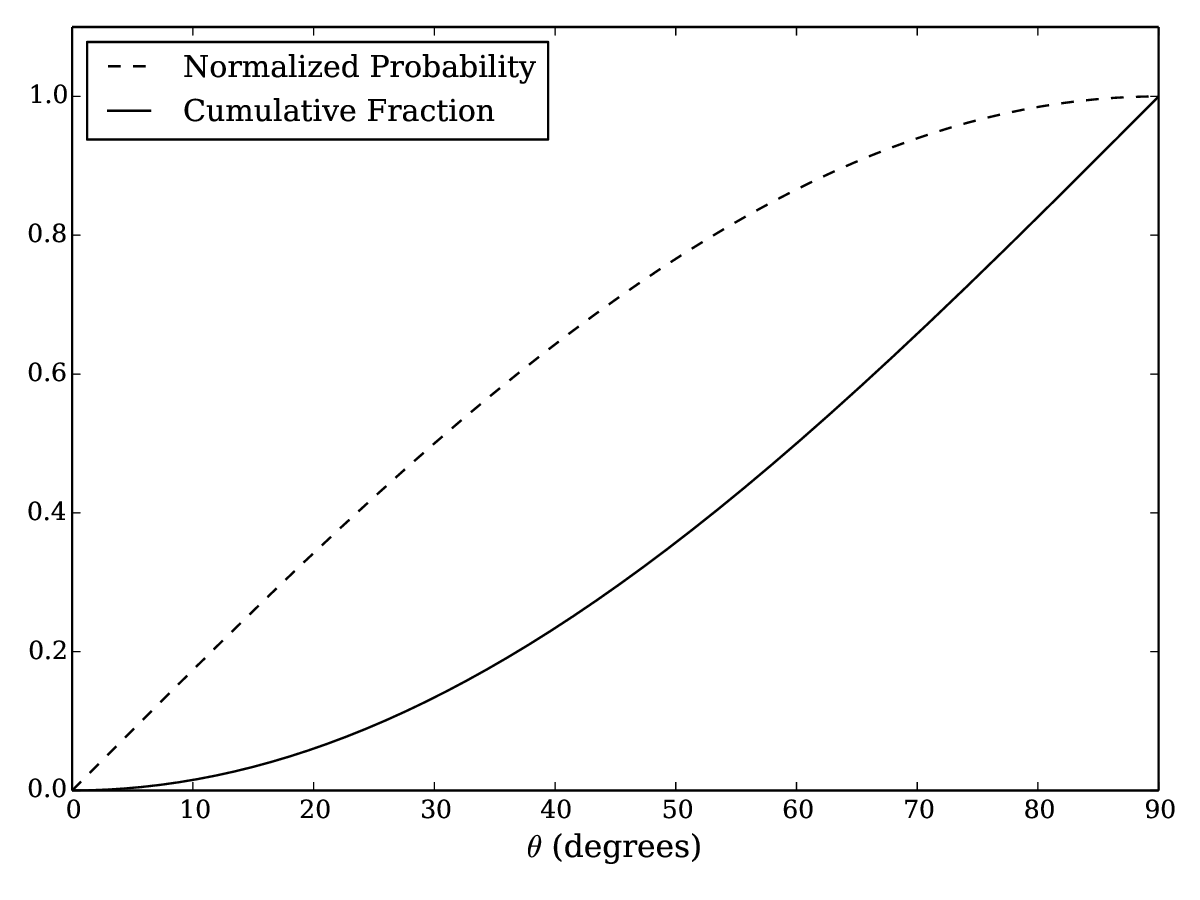}
  \end{center}
  \caption{\textcolor{black}{The dashed curve and solid curve show normalized probability of collision angle $\theta$ ($0^{\circ}\,\leq \theta \,\leq \,90^{\circ}$) and its cumulative fraction, respectively, when we assume that the motion of the colliding clouds are random. $\theta = 0^{\circ}$ means that the collision angle is parallel to the line of sight. }}\label{fig:CCC_prob}
\end{figure}

\textcolor{black}{
Figure\,\ref{fig:CCC_prob} shows normalized probability of collision angle $\theta$ ($0^{\circ}\,\leq \theta \,\leq \,90^{\circ}$) and its cumulative fraction, when we assume that the motion of the colliding clouds are random.
The normalized probability is relative to sin$\theta$, and thus, its cumulative fraction is $1-$cos$\theta$. 
}

\end{document}